\documentclass[11pt]{article}
\pdfoutput=1

\usepackage{standalone}
\usepackage{musixtex}

\usepackage{euscript}
\usepackage{amsfonts}
\usepackage{amsbsy}
\usepackage{epsfig}
\usepackage{amsthm}
\usepackage{amscd}
\usepackage{amstext}
\usepackage{verbatim}
\usepackage{cancel}
\usepackage{capt-of}
\usepackage{empheq}

\usepackage[nosort]{cite}
\usepackage{bm}
\usepackage{authblk}
\usepackage{esint}
\usepackage[T1]{fontenc}
\usepackage{mathdots}
\usepackage[centertableaux]{ytableau}

\usepackage[utf8]{inputenc}
\usepackage{graphicx}
\usepackage{physics}
\usepackage{mathtools}
\usepackage{bbold}

\usepackage{setspace}
\usepackage{colortbl}
\usepackage{xcolor}

\usepackage{amsmath}
\numberwithin{equation}{section}
\makeatletter
\renewcommand*\env@matrix[1][\arraystretch]{\edef\arraystretch{#1}\hskip -\arraycolsep
  \let\@ifnextchar\new@ifnextchar
  \array{*\c@MaxMatrixCols c}}
\makeatother
\usepackage{amssymb}
\usepackage{mathrsfs}
\usepackage{mathtools}
\usepackage{physics}
\usepackage{booktabs}
\usepackage{bbm}
\usepackage{float}

\usepackage{pgfplots}
\pgfplotsset{compat=1.15}
\usepackage{tikz}
\usetikzlibrary{external}
\usetikzlibrary{arrows}

\usepackage{subcaption}
\usepackage{graphicx}
\usepackage{mathtools}
\usepackage[many]{tcolorbox}
\tcbset{shield externalize}
\usepackage{xcolor, colortbl}
\usepackage[a4paper]{geometry}
\usepackage{amsthm}
\theoremstyle{definition}

\usepackage[hidelinks,linktocpage]{hyperref}
\hypersetup{
    colorlinks=true,
    linkcolor=blue,
    citecolor=blue,
    }

\usepackage{mathrsfs}

\renewcommand{\arraystretch}{2} 
\def\ben{\begin{equation}}
\def\een{\end{equation}}
\def\half{{\textstyle{\frac{1}{2}}}}
\let\a=\alpha    
   \let\k=\kappa

\let\w=\omega

\let\pa=\partial
\def\be{\begin{equation}}
\def\ee{\end{equation}}
\def\beq{\begin{equation}}
\def\eeq{\end{equation}}
\def\ba{\begin{array}}
\def\ea{\end{array}}

\def\dalemb#1#2{{\vbox{\hrule height .#2pt
       \hbox{\vrule width.#2pt height#1pt \kern#1pt
               \vrule width.#2pt}
       \hrule height.#2pt}}}

\newcommand{\bea}{\begin{eqnarray}}
\newcommand{\eea}{\end{eqnarray}}

\def\vep{{\varepsilon}}

\newcommand{\e}{\mathrm e}

\makeatletter
\newcommand*\bigcdot{\mathpalette\bigcdot@{.5}}
\newcommand*\bigcdot@[2]{\mathbin{\vcenter{\hbox{\scalebox{#2}{$\m@th#1\bullet$}}}}}
\makeatother

\def\Z{{{\mathbb Z}}}

\def\ocal{{\mathcal{O}}}

\title{Can one hear the shape of a black hole singularity?}
\author{Sean A. Hartnoll$^{\flat}$ and Alexander Zhiboedov$^\sharp$}
\affil{\it $^{\flat}$ Department of Applied Mathematics and Theoretical Physics, \\
\it University of Cambridge, Cambridge CB3 0WA, UK
}
\affil{\it $^{\sharp}$
CERN, Theoretical Physics Department, Geneva, Switzerland
}

\date{}

\begin{document}

\begin{flushleft}
\hfill \parbox[c]{40mm}{CERN-TH-2026-217}
\end{flushleft}

\begingroup
\let\newpage\relax
\maketitle
\endgroup

\begin{abstract}
We compute the large overtone quasinormal frequencies $\omega_n$ of asymptotically AdS black holes with interior near-singularity Kasner epochs. We show that the Kasner exponents can be extracted cleanly from the large $n$ behaviour of the  dispersion $\frac{\pa \omega_n}{\pa k^2}$ at $k=0$, where $k$ is a boundary momentum. We show that Kasner transitions, as occur in BKL near-singularity chaos, are audible in the quasinormal frequencies and reveal a map between overtone number $n$ and proper time to the singularity. Our results are obtained using a WKB analysis that isolates non-analytic exponents, determined by the near-singularity spacetime, from non-universal data such as the optical distance from the boundary to the singularity. We check our formulae in several numerical models.

\end{abstract}
\newpage

\tableofcontents

\newpage

\section{Introduction}

Black hole singularities are a robust consequence of classical gravity \cite{Senovilla:2014gza}, indicating the breakdown of the classical theory. The AdS/CFT correspondence \cite{Maldacena:1997re} is a non-perturbative framework for quantum gravity and hence there is a longstanding effort to identify signatures of the singularity in boundary field theory observables \cite{Fidkowski:2003nf}. Because of the smoothness properties of black hole backgrounds, it is possible to access the interior geometry by analytic continuation of exterior observables. An important early use of this fact was the demonstration that the asymptotic quasinormal mode spectrum can be obtained analytically from the interior geometry \cite{Motl:2003cd}, extended to AdS black holes in \cite{Cardoso:2004up, Natario:2004jd}. Quasinormal modes are the fundamental dissipative excitations of the black hole exterior \cite{Berti:2009kk} and will play a central role in our discussions below.

The signature that appears most directly in boundary observables is the (complex) boundary time it takes a null geodesic to reach down to the singularity. This quantity was firstly shown to appear in the heavy-probe limit of boundary two-point functions, after a suitable analytic continuation in the boundary time \cite{Fidkowski:2003nf}. A more direct imprint of so-called `bouncing geodesics' was found in \cite{Festuccia:2005pi,festuccia2007black}, where it was shown that the time to the singularity controls the behaviour of the two-point function at large imaginary frequencies, as well as the behaviour of the quasinormal modes $\omega_n$ in the large-overtone limit $n \gg 1$. More recently, following the OPE analysis of \cite{Ceplak:2024bja}, the same geodesics were shown to control a non-perturbative correction to the retarded two-point function at large real frequencies \cite{Afkhami-Jeddi:2025wra}. Further studies of the signature of the black hole singularity in the thermal two-point function include \cite{Dodelson:2025jff,Jia:2025jbi,AliAhmad:2026wem,Jia:2026pmv,Grozdanov:2026cut,Jia:2026ryl,Arnaudo:2026tcy,Buric:2026qsp,Arnaudo:2026axe,Barrat:2026jfg}.

The geodesic time to the singularity depends upon the entire spacetime the geodesic traverses. It was explained in \cite{Frenkel:2020ysx} that the local `shape' of spacetime close to the singularity, in contrast, was captured by non-analytic corrections to the geodesic length and time, and hence to the two-point function in the heavy-probe limit. Non-analyticities are universal in the renormalisation-group sense of being intrinsic to the near-singularity spacetime.
Relatedly, it has been known for some time that
non-analytic terms in the large-overtone expansion of quasinormal modes arise from corrections to the effective Schr\"odinger potential in the near-singularity region \cite{MaassenvandenBrink:2003as,Musiri:2003bv,Musiri:2005ev,Dodelson:2023vrw,Jia:2024zes,Giombi:2026kdz}. In this paper, we extend these latter calculations to a rather general class of nonzero-temperature black hole backgrounds. We argue that non-analytic terms in the large-overtone expansion of quasinormal modes provide a crisp probe of the structure of spacetime close to the singularity.

An important class of near-singularity geometries is provided by spatially homogeneous Kasner spacetimes, which were early exact solutions to Einstein's equations \cite{Kasner:1921zz}.  Let $-\tau$ be the proper time to the singularity (at $\tau = 0$) and work in $d+2$ spacetime dimensions. The Kasner metric is then
\be\label{eq:genK}
ds^2 = - d\tau^2 + c_t^2 \tau^{2 p_t} dt^2 + \sum_{i=1}^d c_i^2 \tau^{2 p_i} dx_i^2 \,.
\ee
Here $\{p_t, p_i\}$ are called the Kasner exponents. We keep the additional constants $\{c_t^2, c_i^2\}$ because the normalisation of the $\{t,x_i\}$ coordinates will be set at the AdS boundary. Relatedly, while the spatial directions $\{t,x_i\}$ are on equal footing close to the singularity, we have singled out $t$ as it will be related to the boundary time. For example, the near-singularity geometry of the planar four-dimensional Schwarzschild-AdS solution has the form (\ref{eq:genK}) with $p_t = -\frac{1}{3}$ and $p_1 = p_2 = \frac{2}{3}$. However, the Schwarzschild interior is highly non-generic and unstable within the phase space of solutions to Einstein's equations, as we will discuss shortly. A simple avatar of this instability is that, remaining within spatial homogeneity, coupling a scalar field to gravity and turning on a boundary source \cite{Frenkel:2020ysx} shifts the Kasner exponents away from the Schwarzschild values \cite{Doroshkevich:1978aq}.

Our first result pertains to asymptotically AdS black holes that are homogeneous along the boundary directions and which tend to geometries of the form (\ref{eq:genK}) near the singularity. We show that the Kasner exponents can be extracted from the large overtone behaviour of the quasinormal frequencies $\omega_n$. In particular, the quasinormal mode dispersion takes the form
\be\label{eq:intro}
\left. \frac{\pa \omega_n}{\pa k^2} \right|_{k=0} = \;\; \frac{d(\alpha)}{n^\alpha} + \ocal\left(\frac{1}{n}\right) \,, \qquad \text{as} \qquad n \to \infty \,.
\ee
Here $k$ is the boundary momentum in, say, the $x_i$ direction and the exponent is given in terms of the Kasner exponents by
\be
\alpha = 2 \frac{1 - p_i}{1 - p_t} \,.
\ee
To avoid cluttering, we will \emph{not} put indices on $k$ or $\alpha$. However, both $k$ and $\alpha$ \emph{are} directional and will generically take different values for the different boundary directions. Thus from the quasinormal modes one can extract all of the $p_i$, and the remaining exponent $p_t$ is fixed by the relation $p_t + \sum_{i=1}^d p_i = 1$ (appropriate for theories of gravity or gravity coupled to scalar fields).

The coefficient $d(\alpha)$ in (\ref{eq:intro}) depends on the complex boundary time to the singularity ${\mathcal Z}_+$, mentioned above and defined in (\ref{eq:cZ}), as well as data $f_o, h_o$ that is equivalent to the constants $c_t,c_1$ in the Kasner metric (\ref{eq:genK}):
\be
\label{eq:introd}
d(\alpha) = \frac{i e^{- i \frac{\pi}{2} \a}}{16 f_o  {\mathcal Z}_+} \frac{\pi^{3/2} \Gamma(\frac{1-\a}{2})}{\Gamma\left(1 - \frac{\a}{2}\right)^3} \left(\frac{2 h_o {\mathcal Z}_+}{\pi}\right)^{\a} \,.
\ee
The $\ocal(\frac{1}{n})$ in (\ref{eq:intro}) indicates that terms analytic in $\frac{1}{n}$ appear in the result. These can be larger than the leading non-analytic term shown, if $\a > 1$, but are easy to remove.

Our derivation of (\ref{eq:intro}) uses the monodromy technique for asymptotically AdS black holes introduced by \cite{Cardoso:2004up, Natario:2004jd}, while the non-analytic correction is computed along the lines of earlier results in e.g.~\cite{MaassenvandenBrink:2003as,Musiri:2003bv,Musiri:2005ev,Dodelson:2023vrw, Afkhami-Jeddi:2025wra}. One innovation here is that we consider cases where the background is not known analytically, highlighting that the asymptotic spectrum only depends on the optical distance to the singularity together with universal data of the near-singularity geometry. 
This step requires assumptions about the geometry of anti-Stokes lines in the WKB analysis. For this reason we have performed numerical checks of our results: Fig.~\ref{fig:approach} verifies the coefficient $d(\alpha)$ in (\ref{eq:intro}) by numerically computing quasinormal modes in several different backgrounds.

For many interesting theories of gravity and supergravity, a terminal Kasner approach to the singularity is not generic. A more generic scenario involves instead an infinite chaotic sequence of so-called Kasner epochs \cite{Damour:2002et}. This was first argued by BKL \cite{Belinsky:1970ew} and by Misner \cite{Misner:1969hg} for four-dimensional vacuum gravity. The full BKL dynamics becomes increasingly inhomogeneous towards a singularity, due to spatial points decoupling. That aspect of the physics is beyond the present paper, we return to it briefly in the final \S\ref{sec:discuss}.  However,
transitions between Kasner epochs also arise within certain spatially homogeneous setups. For example, Misner's `mixmaster' chaos emerges explicitly in AdS black hole interiors in a theory of gravity coupled to massive vector fields \cite{DeClerck:2023fax, Caceres:2026mug}. We will focus on an even simpler background that exhibits a single Kasner transition in the interior \cite{Arean:2024pzo}. Numerical and analytical computations of the quasinormal spectrum in this background both lead to our second result, Fig.~\ref{fig:transition}. This figure shows a transition in the exponent $\alpha$ of (\ref{eq:intro}) as a function of the mode number $n$. 
That is to say, interior Kasner transitions are also clearly audible by listening carefully to the external quasinormal ringdown of a black hole.

In BKL dynamics, the $m$th Kasner transition occurs at a particular interior time $\tau_{(m)}$ and is 
audible for modes clustered around a particular mode number $n_{(m)}$. The relation between these two quantities is expressed most cleanly in terms of an interior Schr\"odinger time $x$, defined in (\ref{eq:xx}) below and with $x \to 0$ at the singularity, for which
\be\label{eq:nx}
n_{(m)} \sim \frac{|{\mathcal Z}_+|}{\pi}\frac{1}{x_{(m)}} \,.
\ee
During the $m$th Kasner epoch, following the $m$th transition, the mode dispersion is (\ref{eq:intro}) with $\a = \a_{(m)}$ determined by the Kasner exponents of the epoch.
In (\ref{eq:nx}) short bulk timescales, close to the singularity, are related to high frequency quasinormal modes. Such relations were called UV/UV in \cite{Festuccia:2005pi}, by analogy to the usual UV/IR correspondence of AdS holography \cite{Susskind:1998dq}. The two connections are compatible because it is the UV of the boundary Hamiltonian that maps to the bulk IR, whereas the high overtone quasinormal modes, that probe the bulk singularity, are the UV of a modified boost generator in the boundary theory \cite{Saad:2018bqo,Chen:2023hra}. The relation (\ref{eq:nx}) gives a boundary field theory signature of interior BKL chaos.
The relation furthermore tells us how far down the quasinormal spectrum we must go to hit the string or Planck scale in the bulk, and therefore where to look for the resolution (or not) of the singularity in field theory.

\section{Explicit Einstein-scalar model}
\label{sec:ads2kasner}

In this section we set the stage by introducing concrete black hole backgrounds that exhibit AdS$_4$ asymptotics and Kasner near-singularity behaviour. We write down the wave equation (\ref{eq:schr}) for a probe scalar field in this background and review how bouncing geodesics produce the leading order two-sided correlator (\ref{eq:g12}). This correlator depends on the optical distance ${\mathcal Z}_+$, introduced in (\ref{eq:cZ}), that will play a central role throughout.

\subsection{From AdS to Kasner}

To start with, we work in four spacetime dimensions.
A simple class of asymptotically AdS$_4$ black holes with a Kasner interior arises in the theory of a scalar field coupled to gravity \cite{Frenkel:2020ysx}. The theory has Lagrangian density
\be\label{eq:ES}
{\mathcal L} = \frac{1}{2\k^2} \left[\left(R + 6 \right) - \frac{1}{2} \left(g^{ab}\pa_a \phi \pa_b \phi -2 \phi^2 \right) \right]\,,
\ee
in units where the AdS radius is one. Taking the mass squared to have the negative value $-2$ is inessential, but is convenient for numerics and consistent with the Breitenlohner-Freedman bound. We will
consider backgrounds where the metric takes the form
\be\label{eq:back}
ds^2 = \frac{1}{r^2} \left(- f(r) e^{-\chi(r)} dt^2 + \frac{dr^2}{f(r)} + dx_1^2 + dx_2^2\right) \,.
\ee
The equations of motion for $f(r), \chi(r)$ and $\phi(r)$ are given in Appendix \ref{ap:cor}. 

\begin{figure}[h]
    \centering
    \includegraphics[width=0.8\linewidth]{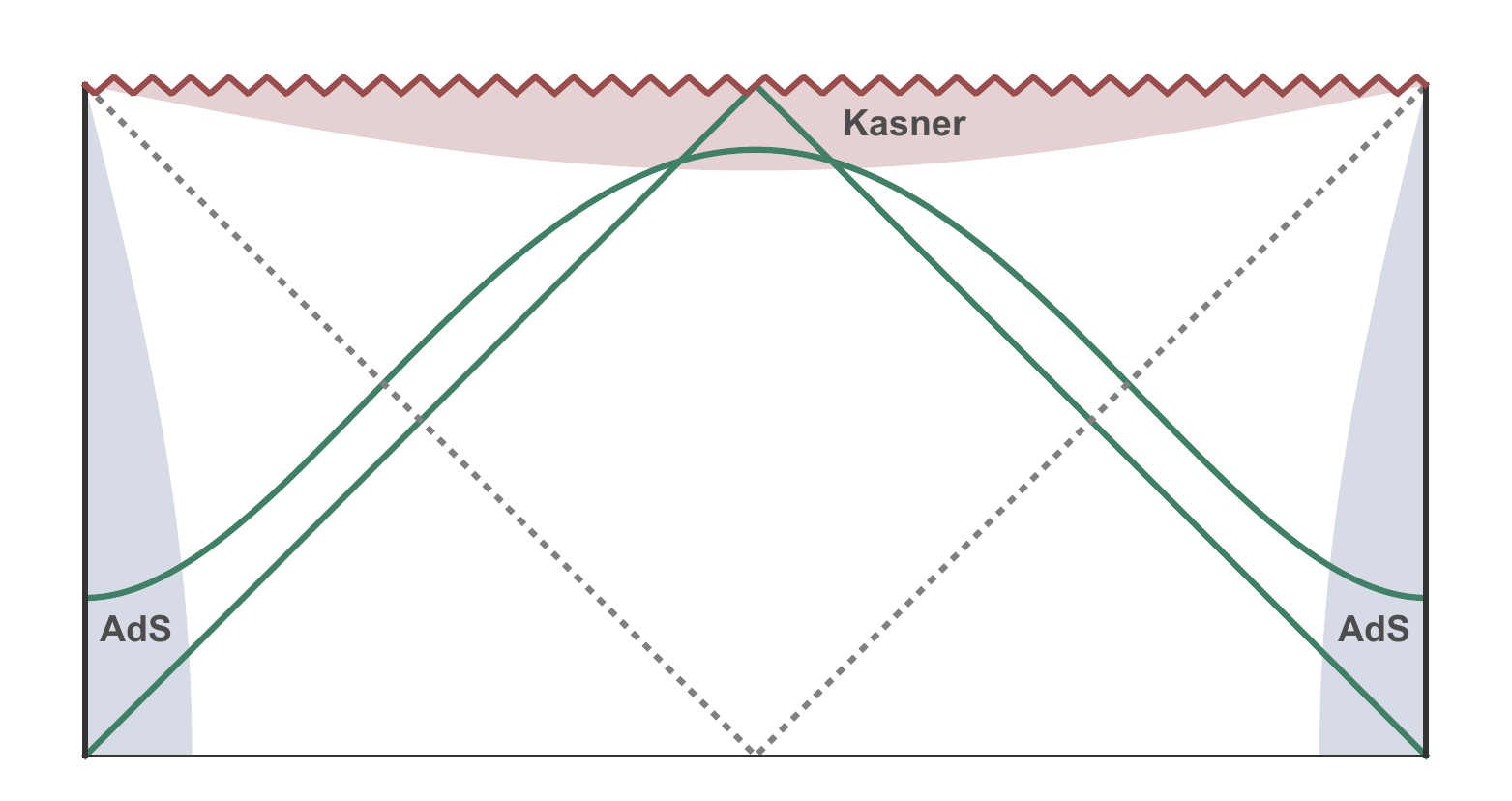}
    \caption{Schematic representation of a spacelike geodesic that travels from one AdS boundary to the other, experiencing a bounce deep in the near-singularity Kasner regime. Also shown is the null limit of spacelike geodesics. The null geodesic goes all the way to the singularity.}
    \label{fig:bounce}
\end{figure}

In the metric (\ref{eq:back}), the AdS boundary is at $r = 0$, where $f \to 1$ and $\chi \to 0$.  A uniform boundary source for the scalar field, $\phi \sim \phi_o r$ as $r \to 0$, corresponds to a relevant deformation of the dual CFT by an operator of dimension $\Delta = 2$. This deformation drives a holographic renormalisation group flow into the bulk spacetime.
The function $f$ changes sign at the black hole horizon, and the singularity is at $r \to \infty$. In that limit one finds a Kasner geometry (\ref{eq:genK}) which, with the metric in the form (\ref{eq:back}), means that in the near-singularity Kasner regime
\be\label{eq:kas} 
f_K = - f_o r^{4/(1- p_t)} \,, \qquad \chi_K = \frac{2(1+ 3 p_t)}{1 - p_t} \log r + \chi_o \,.
\ee
Here $f_o > 0$ and $\chi_o$ are constants and, more importantly, $p_t$ is the Kasner exponent in (\ref{eq:genK}). The rotational symmetry of (\ref{eq:back}) and the constraint that the sum of the Kasner exponents equals one fix the other two exponents to be $p_1 = p_2 = \frac{1 - p_t}{2}$. It was shown in \cite{Frenkel:2020ysx} that, in the theory (\ref{eq:ES}), the exponent $p_t$ varies as a function of the boundary source $\phi_o$ and is found to take values in the range $-\frac{1}{3} \leq p_t < 0$. The fact that $p_t$ is negative means that in this theory the Einstein-Rosen bridge (via $g_{tt}$) always grows towards the singularity. Recall that $p_t = -\frac{1}{3}$ is the value for the Schwarzschild black hole, in the absence of the scalar field. For example, the Ricci scalar in the Kasner regime is $R_K = 2 f_o \frac{3 p_t + 1}{p_t-1} r^{4/(1- p_t)}$. A schematic Penrose diagram of the geometry is shown in Fig.~\ref{fig:bounce}.

\subsection{Scalar field probes and the Kasner potential}

Throughout this paper we will be interested in the quasinormal modes of a probe scalar field $\Phi$. The field obeys the wave equation
\be\label{eq:wave}
\nabla^2 \Phi = m^2 \Phi \,.
\ee
This scalar field is unrelated to the field $\phi$ sourcing the background. Setting
\be
\Phi = r e^{- i \omega t + i k x_1} \Psi(r) \,,
\ee
the wave equation (\ref{eq:wave}) can be written in the Schr\"odinger form
\be\label{eq:schr}
- h \frac{d}{dr}\left(h \frac{d\Psi}{dr} \right) = \left(\omega^2 - V\right) \Psi \,,
\ee
where
\be\label{eq:V}
h \equiv e^{-\chi/2} f \,, \qquad V \equiv \frac{f e^{-\chi}}{r^2}\left(r^2 k^2 + m^2\right) - r h \left(\frac{h}{r^2}\right)' \,.
\ee
Note that $h^2 = - g_{tt}/g_{rr}$.

The physics we are after is due to non-analytic terms in the potential $V$ that arise in the large $r$ Kasner regime. The asymptotic behaviour of the potential is controlled by the near-singularity Kasner exponents, but also by certain non-universal corrections to the Kasner scaling in (\ref{eq:kas}). In Appendix \ref{ap:cor} we give the leading corrections to the Kasner scaling in the Einstein-scalar model. At large $r$ the functions (\ref{eq:V}), appearing in the wave equation, are found to behave as
\be\label{eq:VK0}
h = - h_o r^3 \left(1 +  \frac{F_h(\log r)}{f_o r^{4/(1 - p_t)}} + \cdots \right)  \,, \; V = - h_o^2 r^4 \left(1 + \frac{r^2 k^2}{f_o r^{4/(1 - p_t)}} + \frac{F_V(\log r)}{f_o r^{4/(1 - p_t)}} + \cdots \right) \,.
\ee
Recall that $-\frac{1}{3} \leq p_t < 0$. Here we set $h_o \equiv f_o e^{-\chi_o/2}$, while $F_h(x)$ and $F_V(x)$ are known functions given in Appendix \ref{ap:cor}. In principle we could deal with these functions in our analysis. However, they are non-universal in the sense that they depend on the bulk theory in which the Kasner geometry emerges. To focus on phenomena that are intrinsic to the near-singularity regime we will consider a nonzero momentum, $k \neq 0$. In (\ref{eq:VK0}) we see that the $k$-dependent correction is stronger than the corrections depending on the non-universal functions, because of the $r^2$ in the numerator. We will show that this $k^2$ term in the potential leads to robust non-analyticities in correlators and quasinormal modes. The subleading corrections in (\ref{eq:VK0}) can be safely neglected for these purposes. That is, we may take in the remainder
\be\label{eq:VK}
h_K = - h_o r^3 \,, \qquad V_K = - h_o^2 r^4 \left(1 + \frac{k^2}{f_o} \frac{1}{r^{2 \alpha}}\right) \,, \qquad \alpha \equiv \frac{1 + p_t}{1 - p_t} = 2 \frac{1 - p_1}{1 - p_t} \,.
\ee
The second expression for $\a$ here is the one that will generalise to higher dimensions and without rotational symmetry.

\subsection{The bouncing geodesic at leading order}
\label{sec:bounce0}

A certain high energy limit of the Green's function of the scalar field (\ref{eq:wave}) can be obtained by considering classical geodesics \cite{Fidkowski:2003nf, Festuccia:2005pi, festuccia2007black, Frenkel:2020ysx}, as we now review. This analysis leads to the optical distance ${\mathcal Z}_+$ in (\ref{eq:cZ}) that will re-appear later in the computation of quasinormal modes.

The potential in (\ref{eq:VK}) is negative everywhere in the Kasner regime. A classical turning point at $r_\star$, with $\omega^2 = V(r_\star)$ in the Kasner geometry, can therefore only exist at imaginary frequencies
\be
\omega = - i E \,.
\ee
At large $E$ the turning point occurs close to the singularity, at large $r$. The first term in $V_K$ in (\ref{eq:VK}) dominates at large $r$ because $\a > 0$. Therefore the large $E$ turning point is
\be\label{eq:rstar0}
r_\star \approx r_{\star 0} \equiv \sqrt{\frac{|E|}{h_o}} \,.
\ee
Solutions at large imaginary frequency correspond to classical spacelike geodesics from one exterior of the black hole to the other, that bounce off the near-singularity region along the way. A schematic illustration of such a geodesic is shown in Fig.~\ref{fig:bounce}.

\begin{figure}[h]
    \centering
    \includegraphics[width=0.7\linewidth]{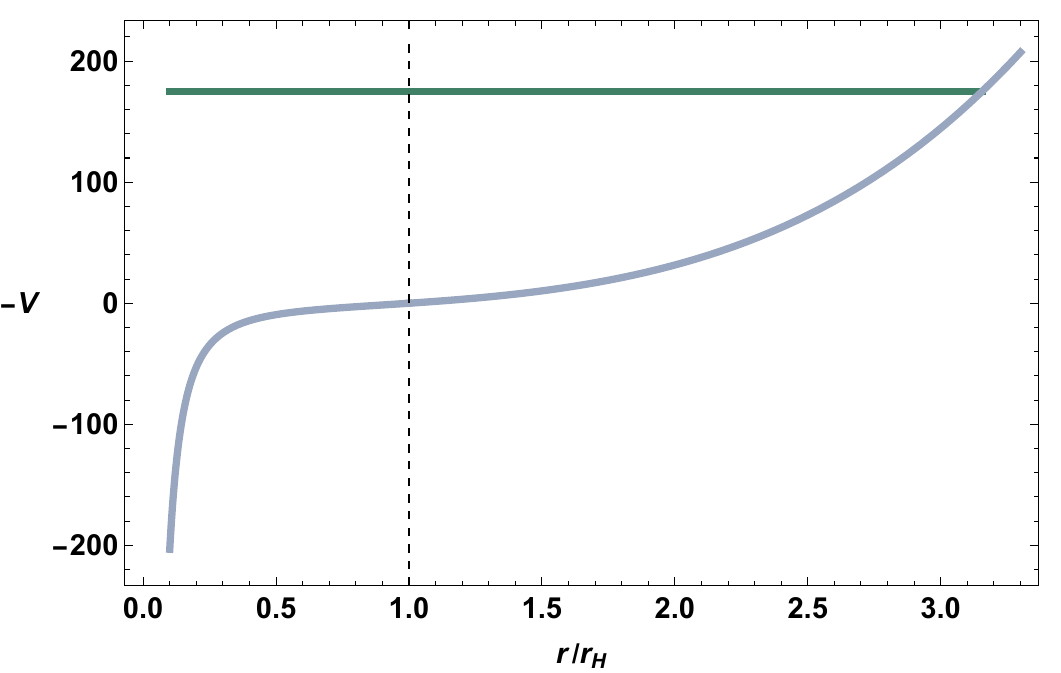}
    \caption{Negative of the potential (\ref{eq:V}) from the boundary at $r=0$ towards the singularity at $r \to \infty$, crossing the horizon at $r = r_{\mathcal{H}}$. Shown with illustrative values $m=0,k=1$ and, in units where the horizon is at $r=1$, solving the equations of motion with data $\phi(1)=\frac{3}{2},\chi(1)\approx 0.358$. A geodesic at constant $E^2=175$ is also shown, illustrating a turning point in the interior region.}
    \label{fig:pot}
\end{figure}

The potential $V$ is positive outside the horizon (so long as $m^2 > -2$), changes sign at the horizon and then grows increasingly negative until hitting the turning point at $r_\star$. This behaviour is illustrated in Fig.~\ref{fig:pot}.
To leading order in the WKB limit, the boundary-to-boundary Green's function is given by the action of a classical geodesic that bounces off the turning point $r_\star$ close to the singularity \cite{Festuccia:2005pi, festuccia2007black},
\be\label{eq:g12a}
G_{12}(- i E) = e^{-2 \,\text{Re}\,S(E)} \,,
\ee
where the factor of 2 captures the ingoing and outgoing parts of the trajectory, and 
\be\label{eq:SE}
S(E) = \int_{r_\text{c}}^{r_\star} \frac{dr}{h} \sqrt{E^2 + V} +\sqrt{2+m^2} \log r_\text{c} \,.
\ee
As usual, $r_\text{c}$ is a near-boundary UV regulator and the final logarithmic term renormalises the boundary divergence (using the facts that $h\approx 1$ and $V \approx \frac{2+m^2}{r^2}$ as $r \to 0$). The zero of $h$ at the horizon leads to an imaginary part in (\ref{eq:SE}), see below. The sign of the imaginary part is determined by a choice of complex $r$ contour. At imaginary frequencies, $\omega = - i E$, the imaginary part of the action cancels between the ingoing and outgoing parts of the geodesic, leading to the real exponent in (\ref{eq:g12a}). Note that the thermal Wightman correlator $G_+(-iE) = e^{- i E/(2T)} G_{12}(-i E)$ does have a complex exponent, in agreement with \cite{Festuccia:2005pi}. Here $T$
is the temperature.

When $E \to \infty$ with $k$ and $m^2$ fixed, the turning point (\ref{eq:rstar0}) reaches the singularity. To leading order in this limit the integral in (\ref{eq:SE}) can be expanded at large $E$ and the limits of integration can safely be taken to zero and infinity, so that
 \be\label{eq:lead}
S(E) \approx E \int_{0}^{\infty} \frac{dr}{h} \,.
\ee
Recalling that $\frac{1}{h} = \sqrt{- \frac{g_{rr}}{g_{tt}}}$, the leading term (\ref{eq:lead}) is $E$ times the boundary time for a null radial geodesic to reach the singularity. The null geodesic is shown in Fig.~\ref{fig:bounce}. This `optical distance' is complex due to $h$ vanishing at the horizon,
\be\label{eq:cZ}
{\mathcal Z}_\pm \equiv \int_{0}^{\infty} \frac{dr}{h} = P \int_{0}^{\infty} \frac{dr}{h} \pm \frac{i}{4 T} \,.
\ee
Here $T$ is the temperature of the horizon \cite{Fidkowski:2003nf,Festuccia:2005pi}. As noted previously, the imaginary parts in (\ref{eq:cZ}) cancel between the two parts of the trajectory contributing to the two-sided correlator (\ref{eq:g12}). To leading order in the large $E$ limit we may therefore write
\be\label{eq:g12}
G_{12}(- i E) \approx e^{-2 E \,\text{Re}\,{\mathcal Z}_+} \,.
\ee
We obtain corrections to this limit in Appendix \ref{ap:two}. The easiest way to do that, it turns out, is to firstly compute the asymptotic quasinormal modes.

While our focus in the remainder will be on quasinormal modes, Appendix \ref{ap:two} shows that the
near-singularity information is equally encoded in
non-analytic corrections to the Green's function at large real $\omega$ or large $E$. Direct computation of these corrections, as in \cite{Afkhami-Jeddi:2025wra}, may provide an interesting alternative perspective on the methods we develop below.

\section{Quasinormal modes}
\label{sec:qnm}

In this section we compute the large overtone quasinormal frequencies of a probe scalar field in a planar AdS black hole background with a Kasner interior. The leading order result (\ref{eq:modes1}) depends only on the optical distance ${\mathcal Z}_+$ to the singularity. The leading non-analytic correction to that result can be extracted directly from the dispersion $\frac{\pa \omega_n}{\pa k^2}$ in (\ref{eq:dwdk}) and encodes the Kasner exponents through the power $\a$. In \S\ref{sec:numericsA} we check our formulae against numerical computations, with good agreement shown in Fig.~\ref{fig:approach}. In \S\ref{sec:general} we establish that our results extend to general Kasner backgrounds, including cases where $g_{tt}$ contracts rather than expands towards the singularity.

\subsection{The large overtone limit}
\label{sec:largen}

\begin{figure}[h]
    \centering
    \includegraphics[width=0.55\linewidth]{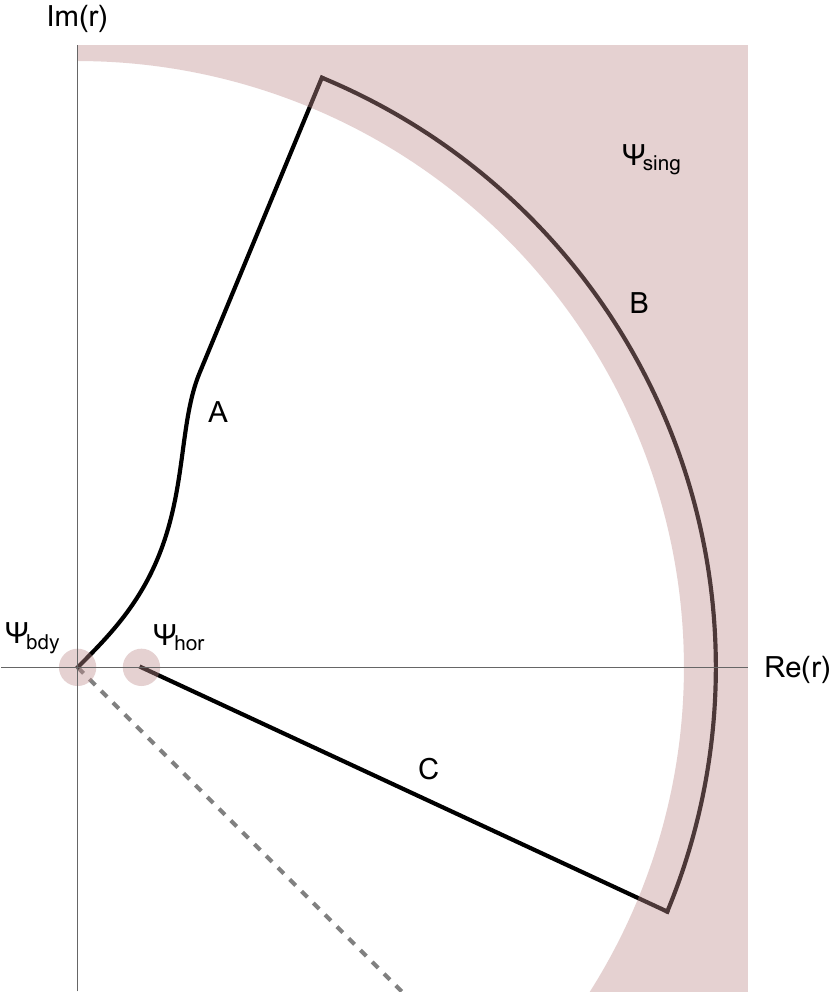}
    \caption{Schematic plot of the complex $r$ contour to be followed in the WKB analysis, see \cite{Cardoso:2004up, Natario:2004jd} for precise contours in the case of Schwarzschild-AdS. The solution $\Psi_\text{bdy}$, that is normalisable at the boundary, and the solution $\Psi_\text{hor}$, that is infalling at the horizon, must be matched along anti-Stokes lines. These are lines where $\text{Im} \left(\omega \int^r \frac{dr}{h} \right)=\text{const}$, so that
    the WKB solution is purely oscillatory. The anti-Stokes lines $A$ and $C$ go to the near-singularity region. They are connected by a path $B$ in the near-singularity regime, leading to a nontrivial monodromy from the solution $\Psi_\text{sing}$ in that region. The asymptotic quasinormal frequencies lie along the dashed line shown.}
    \label{fig:rplane}
\end{figure}

The quasinormal modes are a highly physical and single-sided characterisation of the exterior dynamics. In this section we apply the WKB method of \cite{Cardoso:2004up, Natario:2004jd}, which built on \cite{Motl:2003cd}, to obtain the large overtone limit of the quasinormal modes in terms of near-singularity data. The logic is as follows: solutions to the wave equation that are infalling at the horizon and normalisable at the boundary are only possible for certain discrete complex frequencies. These frequencies can be found by matching the WKB waveform across different regimes. However, because the frequencies are complex this matching must be performed along anti-Stokes lines in the complex $r$ plane, where the WKB waveforms are purely oscillatory. To get from the AdS boundary to the horizon along anti-Stokes lines one is forced to go via the singularity. The calculation below will assume that the topology of the anti-Stokes lines is similar to that of Schwarzschild-AdS. We do not know this because the metric functions are only known numerically at real $r$. We will check our results numerically in \S\ref{sec:numericsA}.
A schematic plot of the contour in the complex $r$ plane is shown in Fig.~\ref{fig:rplane}. This figure may be helpful to follow the steps below.

The normalisable solution to the wave equation (\ref{eq:schr}) near the AdS boundary at $r=0$ is (again using $h\approx 1$ and $V \approx \frac{2+m^2}{r^2}$)
\be\label{eq:nearbdy}
\Psi_\text{bdy} = \sqrt{r} J_\nu\left(r \sqrt{\omega^2 - k^2}\right) \sim e^{i \omega r} + e^{i\pi(\nu + \frac{1}{2})} e^{- i \omega r} \,,
\ee
with $\nu \equiv \sqrt{\frac{9}{4} + m^2}$ and in the final step we have expanded at large $|\omega r|$ to leading exponential order. In this step we have dropped subleading $k^2/\omega$, and higher order, terms. The $k$-dependence originating from these terms can only produce analytic corrections to our result (\ref{eq:dwdk}) below. The near-boundary solution (\ref{eq:nearbdy}) can now be identified, using that $h\approx 1$ at small $r$, with the WKB solution
\be\label{eq:wkb}
\Psi_\text{bdy} = e^{i \omega \int_0^r \frac{dr}{h}} + e^{i\pi(\nu + \frac{1}{2})} e^{- i \omega \int^r_0 \frac{dr}{h}} \,.
\ee
We are looking for complex $\omega$. To extend (\ref{eq:wkb}) away from the boundary we should follow an anti-Stokes line with $\text{Im} \left(\omega \int^r_0 \frac{dr}{h} \right)=0$, where the solution remains oscillatory. We assume that, as is the case for Schwarzschild-AdS, one of these lines can be followed to the near-singularity large $r$ region. This is contour $A$ in Fig.~\ref{fig:rplane}. At small $r$ the anti-Stokes line has $\arg r = -\arg \omega$. 

Near the horizon the wave must be purely infalling, so that
\be\label{eq:hor}
\Psi_\text{hor} = e^{i \omega \int^r \frac{dr}{h}} \,.
\ee
We assume that, as is again the case for Schwarzschild-AdS, there is an anti-Stokes line that allows (\ref{eq:hor}) to be followed to large $r$. This is contour $C$ in Fig.~\ref{fig:rplane}. Along this contour there is no mixing between positive and negative frequency modes. At large $r$ the
wave equation can be solved in terms of Hankel functions, using (\ref{eq:VK}). The infalling solution (\ref{eq:hor}) matches onto
\be\label{eq:sing1}
\Psi_\text{sing} = \frac{1}{r}H_0^{(1)}\left(\frac{\omega}{2 h_o r^2}\right) \sim e^{i \omega \int^r \frac{dr}{h}} \,.
\ee
Here we used that $h \approx - h_o r^3$ in the Kasner regime (\ref{eq:VK}).

There are 4 anti-Stokes lines towards infinity, at $\arg r = \frac{1}{2}\arg \omega + n \frac{\pi}{2}$, with $n=0,1,2,3$. In order to match (\ref{eq:sing1}) with (\ref{eq:wkb}) we must follow the solution through an arc of angle $\frac{\pi}{2}$, to get from the anti-Stokes line $C$ to the anti-Stokes line $A$ in Fig.~\ref{fig:rplane}. This rotation is shown as contour $B$ in Fig.~\ref{fig:rplane}. Using the monodromy properties of Hankel functions we write the solution as
\be
r \Psi_\text{sing} = H_0^{(1)}\left(\frac{\omega}{2 h_o r^2}\right)  = H_0^{(2)}\left(\frac{\omega}{2 h_o (e^{- i \frac{\pi}{2}}r)^2}\right) + 2 H_0^{(1)}\left(\frac{\omega}{2 h_o (e^{- i \frac{\pi}{2}}r)^2}\right) \,. \label{eq:monod}
\ee
The final expression here is written as an oscillatory function along contour $A$ in Fig.~\ref{fig:rplane}, which at large $r$ has $r_A = e^{i \frac{\pi}{2}} r_C$. Only keeping track of the relative magnitude of the leading exponential terms, we expand the Hankel functions and then extend along the $A$ contour to write
\be
\Psi_\text{sing} \sim  e^{i \frac{\pi}{2}}e^{i \omega \int_\infty^r \frac{dr}{h}} + 2 e^{-i \omega \int_\infty^r \frac{dr}{h}} =  e^{i \frac{\pi}{2}}e^{-i \omega {\mathcal Z}_+}e^{i \omega \int_0^r \frac{dr}{h}} + 2 e^{i \omega {\mathcal Z}_+}e^{-i \omega \int_0^r \frac{dr}{h}}  \,. \label{eq:m2}
\ee
Recall that ${\mathcal Z}_+$ was defined in (\ref{eq:cZ}).
In fact, the extension in (\ref{eq:m2}) produces the integral $\int_0^\infty \frac{dr}{h}$ along the anti-Stokes line $A$ in Fig.~\ref{fig:rplane}. To set this integral equal to ${\mathcal Z}_+$ we have rotated the integration contour to the real $r$ axis. This is possible because the asymptotic behaviour of $h$ ensures that there is no contribution to the integral from the arcs at zero or infinity. To make this contour deformation we must further assume, as is the case for Schwarzschild-AdS, that there are no singularities of $1/h$ in between the anti-Stokes line and the real axis. As already noted, we perform numerical computations of quasinormal modes in \S\ref{sec:numericsA} to verify our results.

Finally, matching (\ref{eq:m2}) with (\ref{eq:wkb}) gives the quasinormal mode quantisation condition
\be\label{eq:cond}
1 + 2 e^{i (2 \omega {\mathcal Z}_+ - \pi \nu)} = 0 \,. 
\ee
The solutions to (\ref{eq:cond}) are the modes
\be\label{eq:modes1}
\omega_{0n} = \frac{\pi}{{\mathcal Z}_+} \left(n + \frac{i}{2 \pi} \log 2 + \frac{1}{2} \left(\nu - 1 \right) \right) \equiv \frac{\pi}{{\mathcal Z}_+} N_n \,.
\ee
The $0$ subscript is because we will shortly obtain a non-analytic correction to this formula. The dimensionless quantity $N_n$ will be useful later.
The expression (\ref{eq:modes1}) agrees with \cite{Cardoso:2004up} for a massless scalar in Schwarzschild-AdS$_4$, which has $\nu = \frac{3}{2}$. It also agrees with the result for Schwarzschild-AdS$_{d+2}$ at general $\nu$ \cite{Dodelson:2023vrw}, suggesting that (\ref{eq:modes1}) holds in general dimensions with $\nu^2 = m^2 + \frac{1}{4} (d+1)^2$. Note that \cite{Cardoso:2004up} uses conventions with the opposite sign of $\omega$ and also that there is another set of modes at $-\omega^*_n$. It is remarkable that the large overtone expansion only depends on details of the metric coefficients through the optical distance ${\mathcal Z}_+$. The rest is fixed by the geometry of the anti-Stokes lines.

\subsection{Non-analytic correction in the large overtone limit}\label{sec:qnm2}

We may now compute a non-analytic, in $n$, correction to the asymptotic quasinormal mode spectrum (\ref{eq:modes1}). The non-analyticity originates from a correction to the monodromy step (\ref{eq:monod}) in the near-singularity region. Similar computations can be found in e.g.~\cite{MaassenvandenBrink:2003as,Musiri:2003bv,Musiri:2005ev,Dodelson:2023vrw} and \cite{Afkhami-Jeddi:2025wra}.

The starting point is the wave equation in the Kasner regime with potential (\ref{eq:VK}). We wish to obtain the shift in the monodromy due to the subleading term in this potential. To keep things general we write the potential as
\be\label{eq:pot3}
V_K = - h_o^2 r^4 (1 + \frac{\vep}{r^{2\a}}) \,,
\ee
with $\vep = \frac{k^2}{f_o}$ in the case of immediate interest. Introduce the variable $x \equiv \omega/(2 h_o r^2)$ and let $\Psi(x) = \sqrt{x} \, \psi(x)$. The wave equation (\ref{eq:schr}), in the Kasner regime, then becomes
\be\label{eq:xeq}
\psi'' + \frac{1}{x} \psi' + \psi = -  \hat\vep \, x^{\a-2} \psi \,, \qquad \hat \vep \equiv \frac{\vep}{4} \left(\frac{2 h_o}{\omega} \right)^{\a} \,.
\ee
The zeroth order solution to (\ref{eq:xeq}), setting $\hat \vep = 0$, is the near-singularity infalling wave (\ref{eq:sing1}).

We now solve (\ref{eq:xeq}) to first order in $\hat \vep \neq 0$ by writing
\be\label{eq:psi}
\psi(x) = H_0^{(1)}(x) + \hat \vep \, \delta \psi(x) \,.
\ee
The zeroth order term sources the perturbation, which is then found using the Green's function for (\ref{eq:xeq}). Thus the solution is expected to involve a product of three Hankel functions, one from the source and two from the Green's function. Indeed, the required solution is
\be
\delta \psi(x) = \frac{i \pi}{4} \left[-H_0^{(1)}(x) \int_x^\infty t^{\a-1} H_0^{(1)}(t) H_0^{(2)}(t) dt + H_0^{(2)}(x) \int_x^\infty t^{\a-1} \left(H_0^{(1)}(t)\right)^2 dt  \right] \,. \label{eq:l1}
\ee
The limits of integration must ensure that the $H_0^{(2)}(x)$ term vanishes at large $x$, so that the perturbation remains infalling. The limits for the $H_0^{(1)}(x)$ term are not fixed, as we are free to shift $\delta \psi(x)$ by any multiple of the zeroth order solution in (\ref{eq:psi}). This ambiguity drops out of the monodromy shift, and
we have made a convenient choice in (\ref{eq:l1}). To compute the monodromy it will be best to work with integrals that run from zero to $x$, as these are easier to rotate in the complex $x$ plane. To this end we write $\int_x^\infty dt = \int_0^\infty dt - \int_0^x dt$ and perform the integrals from zero to infinity, to obtain
\begin{align}
\delta \psi(x) & = \frac{i \pi}{4} \Bigg[
- \frac{e^{\frac{i \pi \a}{2}} \Gamma\left(\frac{\a}{2}\right)^3}{\pi^{\frac{3}{2}} \Gamma\left(\frac{1 + \a}{2}\right)} H_0^{(2)}(x) - \frac{\Gamma\left(\frac{\a}{2}\right)^3 \Gamma\left(\frac{1 - \a}{2}\right)}{\pi^{5/2}} H_0^{(1)}(x)  \nonumber \\
& \qquad\qquad + \int_0^x t^{\a-1} H_0^{(1)}(t) \left( H_0^{(1)}(x) H_0^{(2)}(t) - H_0^{(2)}(x) H_0^{(1)}(t)\right)dt  \Bigg] \,. \label{eq:l2}
\end{align}
The integrals converge if $0 < \a < 1$. The final result (\ref{eq:wfull}) can be analytically continued outside of this range. In a closely related computation in Appendix \ref{app:analytic-alpha} we see explicitly how this analytic continuation is performed.

To compute the correction to the monodromy (\ref{eq:monod}) we express $\delta\psi(x)$ as a function of $e^{i \pi}x$. We may do this in (\ref{eq:l2}) as follows (this paragraph outlines the steps, more details can be found in Appendix \ref{app:first-order-monodromy}). Firstly, by writing $H^{(1)}_0(x) = 2 H^{(1)}_0(e^{i\pi}x)+H^{(2)}_0(e^{i\pi}x)$ and $H^{(2)}_0(x) = - H^{(1)}_0(e^{i\pi}x)$. Secondly, we change integration variables by setting $t = e^{-i \pi}s$, which makes the upper limit of integration $e^{i \pi} x$. Using the transformations just given, we rotate the Hankel functions in the integrand to the real $s$ axis, e.g.~$H^{(2)}_0(e^{- i \pi}s) = - H^{(1)}_0(s)$.
To obtain a correction to (\ref{eq:monod}) we expand at large $x$ by sending the upper limit of integration to infinity. The resulting integrals can be performed and give, using gamma function reflection formulae to simplify the expression,
\be\label{eq:shift}
\delta \psi(x) \sim \frac{e^{- i \frac{\pi}{2} \a} \pi^{3/2} \Gamma(\frac{1-\a}{2})}{\Gamma\left(1 - \frac{\a}{2} \right)^3}  H_0^{(1)}(e^{i\pi}x)  \, . 
\ee
The absence of an $ H_0^{(2)}(e^{i\pi}x)$ term in (\ref{eq:shift}) is due to the choice of integration limits in (\ref{eq:l1}).

The Hankel function in (\ref{eq:shift}) produces a shift in one of the coefficients on the right hand side of (\ref{eq:monod}): $2 \to 2 + \hat \vep X \approx 2 e^{\frac{\hat \vep}{2} X}$, where $X$ is the prefactor in (\ref{eq:shift}). This shift is then incorporated into the quantisation condition (\ref{eq:cond}) as $2 \to 2 e^{\frac{\hat \vep}{2} X}$. Solving the shifted quantisation condition gives the corrected quasinormal modes
\be\label{eq:wfull}
\omega_n = \omega_{0n} + \vep \frac{i e^{- i \frac{\pi}{2} \a} }{16 {\mathcal Z}_+} \frac{\pi^{3/2} \Gamma(\frac{1-\a}{2})}{\Gamma\left(1 - \frac{\a}{2}\right)^3} \left(\frac{2 h_o}{\omega_{0n}}\right)^{\a} + \cdots \,.
\ee
The $\cdots$ include analytic corrections in $\frac{1}{\omega_{0n}}$ as well as higher order non-analytic corrections. We comment further on the analytic terms shortly. Restoring $\vep = \frac{k^2}{f_o}$ for the case of interest, we can cleanly extract the large $n$ non-analytic correction in (\ref{eq:wfull}) by taking the derivative. Specifically,
\be\label{eq:dwdk}
\left.\frac{\pa \omega_n}{\pa k^2}\right|_{k=0} = \;\;\;  \frac{i e^{- i \frac{\pi}{2} \a}}{16 f_o  {\mathcal Z}_+} \frac{\pi^{3/2} \Gamma(\frac{1-\a}{2})}{\Gamma\left(1 - \frac{\a}{2}\right)^3} \left(\frac{2 h_o {\mathcal Z}_+}{\pi}\right)^{\a} \frac{1}{n^{\a}} + \cdots \,.
\ee
By listening to the high ringdown overtones, therefore, we can indeed hear the near-singularity Kasner exponent. 
This is our first result, advertised in (\ref{eq:intro}) in the introduction.

In (\ref{eq:dwdk}) we have expanded $\omega_{0n} \approx \frac{\pi n}{{\mathcal Z}_+}$ to leading order at large $n$. In our numerical discussions below we will often keep the resummed expression obtained in (\ref{eq:wfull}). That is to say, we can upgrade (\ref{eq:dwdk}) by replacing $n \to N_n$, with $N_n$ defined in (\ref{eq:modes1}).

The mode dispersion (\ref{eq:dwdk}) is an especially clean observable, but there are also $k$-independent corrections to the quasinormal modes that depend on the Kasner exponent $p_t$. For example, there is a $k$-independent perturbation to the near-singularity potential for Schwarzschild-AdS$_4$ of the form (\ref{eq:pot3}), now with $\vep = (m^2 + \frac{9}{5})/f_o$. The other parameters for this perturbation are $f_o = h_o = \left(\frac{4}{3} \pi T\right)^3$, $\a = \frac{3}{2} = \frac{2}{1 - p_t}$ and ${\mathcal Z}_+ = e^{i \frac{\pi}{3}}/(2 \sqrt{3} T)$.
With these parameters, the correction to the quasinormal frequencies in (\ref{eq:wfull}) agrees exactly with the expression given in Appendix F of \cite{Dodelson:2023vrw}, after complex conjugating the full expression to match conventions. We note in Appendix \ref{ap:cor} that the shift by $\frac{9}{5}$ in the expression for $\vep$ here is due to the non-universal corrections contained in the functions $F_h$ and $F_V$ in (\ref{eq:VK0}). More general Kasner singularities in the Einstein-scalar theory have the background scalar field turned on, leading to logarithmic terms in the functions $F_h$ and $F_V$. These logarithms lead to additional $n^{-2/(1 - p_t)}\log^2n$ and $n^{-2/(1 - p_t)}\log n$ corrections to the asymptotic quasinormal modes. While these corrections are more complicated, and we will not consider them in any detail, they do reveal the near-singularity Kasner exponent $p_t$.

Another way to extract a universal term, intrinsic to the singularity, is to differentiate with respect to the mass squared rather than the momentum, $\frac{\pa \omega_n}{\pa m^2}$. This will pick out a correction to the potential of the form (\ref{eq:pot3}) with $\alpha = \frac{2}{1 - p_t}$, generalising the example of Schwarzschild-AdS$_4$ discussed in the previous paragraph and bypassing the logarithmic terms. Probing the singularity through the mass dependence of external observables has been considered previously in \cite{Grinberg:2020fdj}.

To finish this discussion, note that in the non-generic case that $\alpha \geq 0$ is an integer, there is a resonance between the analytic and non-analytic terms. The resonance causes (\ref{eq:wfull}) to be incorrect as written. There are several cases to consider. For $\alpha = 2 j +1$ the prefactor in (\ref{eq:wfull}) diverges. The expectation here is that the analytic divergent part will be renormalised away while the finite part of \eqref{eq:wfull} will correctly capture the non-analytic behavior $\frac{\log n}{n^{2j+1}}$. For $\alpha = 2j>0$, the dispersion vanishes and we cannot use the simple diagnostic considered here to capture the geometry near the singularity. The interesting limiting case of $\alpha=0$ describes the near-singularity geometry of a BTZ black hole and a black hole in the self-dual linear axion model discussed in \cite{Grozdanov:2026cut}. The expectation here is that there will now be a $k$-dependent term in the leading order expression (\ref{eq:modes1}), that must be computed by redoing the steps in \S\ref{sec:largen}. The presence of such terms tells us about near-singularity Kasner exponents (i.e.~$p_i=1$) even while there are no bouncing geodesics in these cases.

\subsection{Numerical checks: Hearing the Kasner geometry}
\label{sec:numericsA}

\begin{figure}[H]
    \centering
    \includegraphics[width=0.85\linewidth]{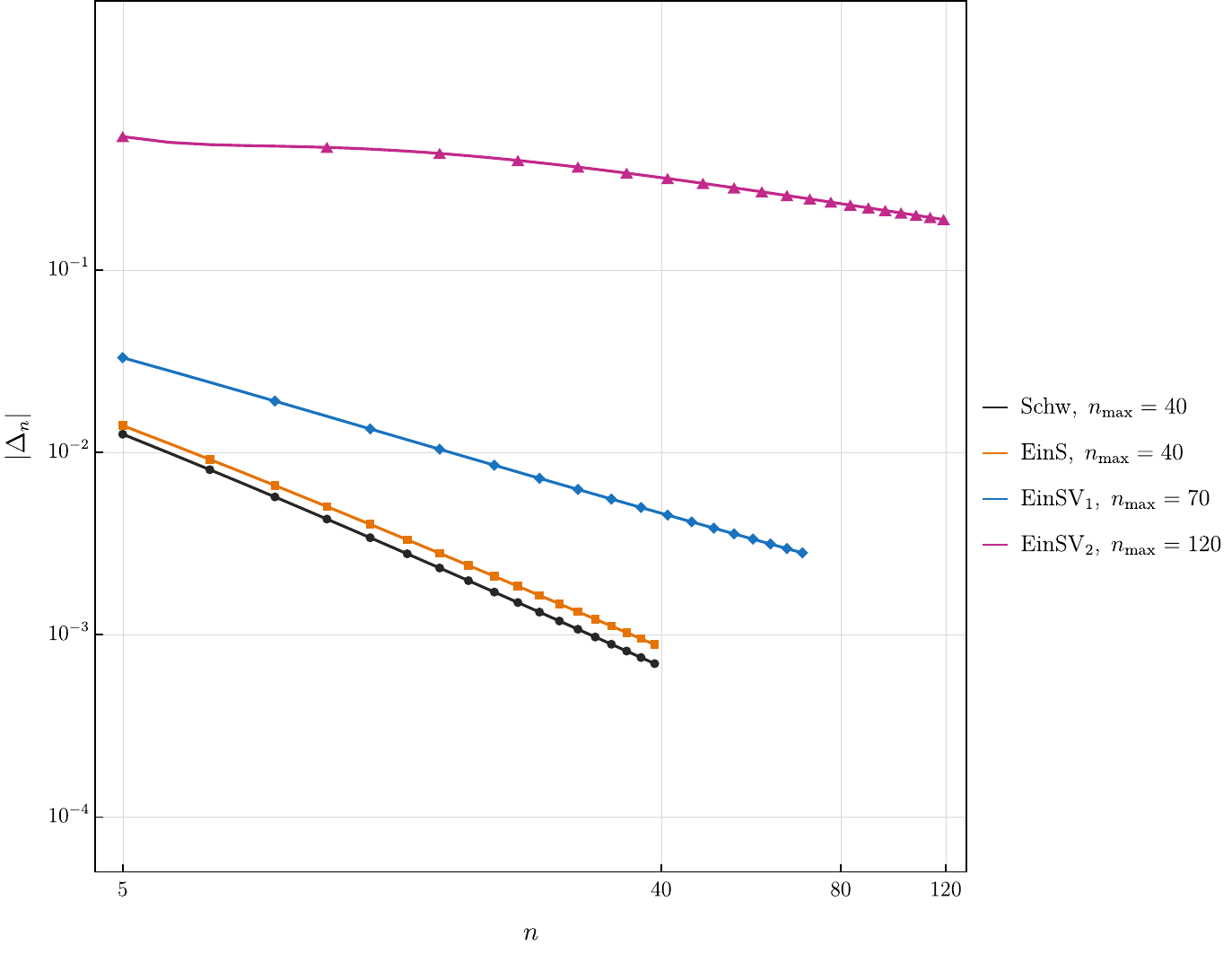}
    \caption{Relative error (\ref{eq:Delta}) between the asymptotic WKB prediction (\ref{eq:dwdk}) and  quasinormal modes computed numerically, as a function of the mode number. The modes have been computed for the four models discussed in this section, summarised in Table \ref{tab:models}. We mark $\sim 20$ equidistant overtones for each curve, starting from $n =5$. In the main text we explain that the slow decay of the \texttt{EinSV}$_2$ curve is a consequence of this background having $\a > 1$.}
    \label{fig:approach}
\end{figure}

In this section we will test the formula (\ref{eq:dwdk}) for $\left. \frac{\pa \omega_n}{\pa k^2}\right|_{k=0}$ with numerical computations of quasinormal modes in several backgrounds. Checks are important because the derivation of (\ref{eq:dwdk}) assumed that the anti-Stokes lines had the same geometry as for the Schwarzschild-AdS background and that there was no obstruction to rotating the integral defining ${\mathcal Z}_+$ from the anti-Stokes line to the real axis. Furthermore, the result was established directly for $0 < \a < 1$ and extension to larger $\a$ required analytic continuation. The main result of this section is Fig.~\ref{fig:approach}, showing that the numerical results are converging towards (\ref{eq:dwdk}) as $n$ is increased.

We test (\ref{eq:dwdk}) on four different geometries.
In each case we choose convenient coordinates for the bulk computations. The first background is Schwarzschild-AdS$_4$. Placing the horizon at $r=1$, the metric has the form (\ref{eq:back}) with $\chi = 0$ and $f = 1 - r^3$. The data required to evaluate (\ref{eq:dwdk}) for this background is shown in the row \texttt{Schw} in Table \ref{tab:models}. The second background is a numerical solution to the Einstein-scalar model (\ref{eq:ES}). We choose a solution with $\phi(1) = \frac{1}{2}$ and $\chi(1) \approx 0.048$ on the horizon at $r=1$ (the choice of $\chi(1)$ is fixed by the requirement that $\chi \to 0$ at the boundary). This leads to a Kasner interior with parameters shown in the row labelled \texttt{EinS} in Table \ref{tab:models}. All of these parameters are close to the Schwarzschild ones.

The remaining two geometries are exact solutions to the following Einstein-scalar model in four dimensions with a potential,
\be\label{eq:ESV}
{\mathcal L} = \frac{1}{2\k^2} \left[R  - \frac{1}{2} g^{ab}\pa_a \phi \pa_b \phi - V(\phi) \right]\,,
\ee
where
\be
V(\phi) = \frac{10 + 9 \lambda}{45} V_o(\phi) - \frac{9 \lambda}{45} V_o(-\phi)   \,, \qquad 
V_o(\phi) \equiv - 16 e^{-{\phi \over 2\sqrt{2}}} - 10 e^{{\phi \over \sqrt{2}}} - e^{-\sqrt{2} \phi} \,.
\ee 
This potential was written down in \cite{Acena:2013jya} to enable an analytic solution to the equations of motion. Having an analytic background makes the numerical computation of quasinormal modes significantly easier. The potential obeys $V(0) = -6, V'(0) = 0$ and $V''(0) = - 2$ and therefore admits unit radius AdS$_4$ asymptotics where the scalar field has a mass squared equal to $-2$, just like in the model (\ref{eq:ES}). We give details of the exact backgrounds in Appendix \ref{ap:exact}. The solutions are asymptotically AdS$_4$, have a regular horizon and a Kasner near-singularity geometry. For two different choices of the parameter $\lambda$, the quantities appearing in the Kasner interior are shown in the rows labelled \texttt{EinSV}$_1$ and \texttt{EinSV}$_2$ in Table \ref{tab:models}. The table also shows the corresponding prediction from (\ref{eq:dwdk}).

\begin{table}[h]
    \centering
    \begin{tabular}{ c  >{$\textstyle}c<{$}  >{$\textstyle}c<{$} >{$\textstyle}c<{$} >{$\textstyle}c<{$} >{$\textstyle}c<{$} >{$\textstyle}c<{$}  }
    \toprule
       model & p_t  & p_1 & f_o & h_o & {\mathcal Z}_+ & \left.\frac{\pa \omega_n}{\pa k^2}\right|_{k=0} \\ \midrule
       \texttt{Schw} & -\frac{1}{3} & \frac{2}{3}  & 1 & 1 & \frac{2 \pi}{3 \sqrt{3}} e^{i \pi/3} & \frac{0.4806 + 0.1288 i}{n^{1/2}} \\ 
       \texttt{EinS} & - 0.3172 & 0.6586 & 1.0479 & 1.0298 & 0.6080 + 1.0295 i & \frac{0.4709 +0.1237 i}{n^{0.5184}} \\ 
    \texttt{EinSV}$_1$ & -\frac{1}{7} & \frac{4}{7} & \frac{24 \sqrt{3}}{37} & \frac{36}{37} & 0.7310 + 0.9489 i & \frac{0.5205 + 0.0862 i}{n^{3/4}} \\
     \texttt{EinSV}$_2$ & \frac{1}{5} & \frac{2}{5} & \frac{4320}{71} & \frac{1440}{71} & 0.4456 + 0.4552 i & \frac{-0.0202 + 0.0082 i}{n^{3/2}}  \\ \bottomrule
    \end{tabular}
    \caption{Quantities characterising the near-singularity Kasner geometry ($p_t,p_1,f_o,h_o$), the optical length ${\mathcal Z}_+$ and the prediction (\ref{eq:dwdk}) for the four models described in the text. Values are shown to four decimal places, but for producing Fig.~\ref{fig:approach} we have them to much higher accuracy.}
    \label{tab:models}
\end{table}

In Appendix \ref{ap:numerical} we describe the numerical computation of $\left.\frac{\pa \omega_n}{\pa k^2}\right|_{k=0}$ for the four backgrounds above. The numerical results recover the predictions in Table \ref{tab:models} to high accuracy, as shown in Fig.~\ref{fig:approach}. To extract the non-analytic terms shown in the table from numerical data, we must remove any
analytic corrections of the form $1/n$. This is especially important whenever $\a > 1$, as a $1/n$ term dominates over $1/n^\alpha$ in this case. The analytic corrections can be calculated explicitly and we have verified that the calculations match the values found in numerics. However, much like ${\mathcal Z}_+$, these corrections do not probe the geometry of the singularity and so we will not discuss them here. Instead, we define a quantity $B_n$ in (\ref{eq:bn}) below that automatically subtracts out the $1/n$ corrections. In this way we access the non-analytic terms while remaining agnostic about the coefficients of the analytic corrections.

Firstly, define for notational convenience
\be
D_n \equiv \left.\frac{\pa \omega_n}{\pa k^2}\right|_{k=0} \,.
\ee
The prediction from \eqref{eq:dwdk}, upgraded with $n \to N_n$, is that at large $n$
\be\label{eq:Aform}
D_n = \frac{d_o}{N_n} + \frac{d(\alpha)}{N_n^\alpha} + \cdots \,.
\ee
Recall from (\ref{eq:modes1}) that $N_n \equiv n+\frac{1}{4}+\frac{i\log 2}{2\pi}$, where in the numerics we fix $\nu$ by taking the probe scalar field to be massless. In (\ref{eq:Aform}) we allowed for an analytic correction, with $d_o$ constant.
The analytic term can be canceled out by defining the discrete derivative
\be\label{eq:bn}
B_n \equiv N_{n+1} D_{n+1}-N_n D_n \,.
\ee
This quantity is a discrete version of $\pa_n (N_n D_n)$.
In particular, if $D_n$ takes the form (\ref{eq:Aform}) then the ratio $B_n/(N_{n+1}^{\,1-\alpha}-N_n^{\,1-\alpha}) = d(\alpha) + \cdots$.
The plots in Fig.~\ref{fig:approach} show that the relative error
\be\label{eq:Delta}
\Delta_n \equiv \frac{1}{d(\alpha)} \frac{B_n}{N_{n+1}^{\,1-\alpha}-N_n^{\,1-\alpha}} - 1 \,,
\ee
\noindent goes to zero as $n$ is increased, if we use the values of $d(\alpha)$ as predicted in Table \ref{tab:models}. Needless to say, a decreasing error for the coefficient $d(\alpha)$ is only possible if the value of $\alpha$ itself in the numerics matches the predictions in Table \ref{tab:models}.

In Fig.~\ref{fig:approach} the \texttt{EinSV}$_2$ case is seen to converge more slowly than the others. The rate of convergence can be understood as follows. The \texttt{EinSV}$_2$ background has $\alpha = \frac{3}{2} > 1$, so the leading correction to (\ref{eq:Aform}) is a further analytic term $\frac{d_1}{N_n^2}$. Including this term in (\ref{eq:Delta}) one finds that to leading order, $\Delta_n = \frac{2 d_1}{d(\alpha)} \frac{1}{\sqrt{N_n}}$. The slope of the \texttt{EinSV}$_2$ curve in the log-log plot Fig.~\ref{fig:approach} is indeed found to be very close to $-\frac{1}{2}$. Of course, a more refined quantity could be defined that subtracts off both the first and second order analytic contributions, leading to faster convergence. As it stands, Fig.~\ref{fig:approach} shows that equation (\ref{eq:dwdk}) is correct for these four backgrounds.

\subsection{General Kasner backgrounds}
\label{sec:general}

In this section we extend the result (\ref{eq:dwdk}) to interior Kasner geometries in higher dimensions and without rotational symmetry.
We wrote the general Kasner metric in (\ref{eq:genK}) in the introduction. The wave equation (\ref{eq:wave}) in this more general Kasner background can again be put in the Schr\"odinger form (\ref{eq:schr}), setting
\be
\Phi = \tau^{-\frac{P}{2}}\Psi(\tau) e^{- i \omega t + i k x_1} \,, \qquad P \equiv \sum_{i=1}^d p_i \,.
\ee
Expanding for small $\tau$, 
\be\label{eq:genV}
h_K = c_t \tau^{p_t} + \cdots  \,, \quad V_K = - c_t^2 \tau^{2 p_t} \left(\frac{P \left(2 - 2 p_t - P\right)}{4 \tau^2} + \frac{k^2}{c_1^2 \tau^{2 p_1}}  +  m^2  + \cdots 
\right)\,.
\ee
Recall that in (\ref{eq:VK0}) the non-universal $+\cdots$ terms competed with the $m^2$ term in the potential. More generally, these non-universal terms can also be equal to or larger than the $k^2$ term as $\tau \to 0$. A universal contribution from the Kasner geometry may nonetheless be picked out through the derivative $\frac{\pa \omega_n}{\pa k^2}$.

The results of the previous sections carry through to the more general geometry (\ref{eq:genK}) most directly
when two conditions are imposed. Both of these conditions pertain for pure gravity and for gravity coupled only to scalar fields, in any dimension \cite{Damour:2002et}. The first is that the Kasner exponents sum to one:
\be\label{eq:sum}
p_t + P = 1 \,.
\ee
Using (\ref{eq:sum}) the numerator of the first term in the potential (\ref{eq:genV}) is manifestly positive, $P\left(2 - 2 p_t - P\right) = (1 - p_t)^2$. Furthermore, we will now see that precisely this numerator is needed to recover the equation of motion (\ref{eq:xeq}) that underpinned the monodromy computation.

The second assumption is that all the Kasner exponents are less than one, $\{p_t, p_i\} < 1$. This second condition implies that the 
Schr\"odinger coordinate
\be\label{eq:xx}
x \equiv \int \frac{d \tau}{h(\tau)} = \frac{1}{(1-p_t) c_t} \tau^{1-p_t} + \cdots \to 0 \qquad \text{as} \qquad \tau \to 0 \,.
\ee
This condition also implies that the first term in the potential (\ref{eq:genV}) dominates and diverges as $\tau \to 0$. In terms of the Schr\"odinger coordinate (\ref{eq:xx}), the potential to leading order becomes
\be\label{eq:V3}
V_K(x) = - \frac{1}{4 x^2}\Big(1 + 4 \vep x^{\a}\Big) + \cdots \,,
\ee
where $\vep = \frac{c_t^2}{c_1^2}(c_t[1-p_t])^{\a-2} k^2$ and now the exponent
\be\label{eq:a12}
\alpha \equiv 2 \frac{1 - p_1}{1-p_t} \,.
\ee
The model in \S\ref{sec:ads2kasner} had $p_t + 2 p_1 = 1$. Setting $\Psi = \sqrt{x}\psi$ and rescaling $x$ by $\omega$, the Schr\"odinger equation with the near-singularity potential (\ref{eq:V3}) becomes the previous equation of motion (\ref{eq:xeq}) with $\hat \vep = \frac{\vep}{\omega^\alpha}$. All of our previous results therefore go through unchanged, now with the more general exponent (\ref{eq:a12}). The condition (\ref{eq:sum}) on the sum of Kasner exponents ensured that the coefficient of the leading $\frac{1}{x^2}$ in the potential (\ref{eq:V3}) is equal to $-\frac{1}{4}$. Otherwise, the near-singularity Hankel functions in the monodromy computation of \S\ref{sec:qnm} have a nonzero order. This fact wouldn't change the exponent, but would change the prefactor of the non-analytic term.

An important aspect of the general case is that the analysis goes through with either sign of $p_t$ (as we have already verified in Table \ref{tab:models}). This means that $g_{tt} \sim \tau^{2 p_t}$ can be expanding or contracting towards the singularity and there will still be a large $E$ geodesic turning point, so long as $p_t < 1$, at
\be\label{eq:taustar}
\tau_\star \approx \left[\frac{c_t(1-p_t)}{2E}\right]^{1/(1-p_t)} \,.
\ee
This occurs because the first term in the potential (\ref{eq:genV}) dominates. This is different from a WKB analysis based on large mass $m$, which is also closely connected to geodesics \cite{Festuccia:2008zx, Hartman:2013qma}. If $p_t$ is positive the $m^2$ term in the potential (\ref{eq:genV}) decreases in magnitude towards the singularity and does not cause a turning point. Asymptotically AdS black holes with $p_t > 0$ Kasner regimes towards the singularity were constructed in \cite{Hartnoll:2020rwq, Hartnoll:2020fhc}. While Maxwell fields are crucial for those backgrounds to exist, their Kasner regimes obey the two conditions we have imposed above. Our results for large overtone quasinormal modes should therefore apply to those spacetimes.

\section{Multiple Kasner epochs}
\label{sec:multi}

The black hole solutions in \S\ref{sec:numericsA} have a single interior Kasner regime. In the introduction we explained that transitions between distinct Kasner epochs are a further important phenomenon in black hole interiors. In this section we show that interior Kasner transitions are also clearly audible in the asymptotic quasinormal mode spectrum, see Fig.~\ref{fig:transition}. Building on this observation, in \S\ref{sec:billiards} we explore possible signatures of interior BKL chaos in the quasinormal mode spectrum.

\subsection{A Kasner transition: numerics}

\begin{figure}[h]
    \centering
    \includegraphics[width=0.95\linewidth]{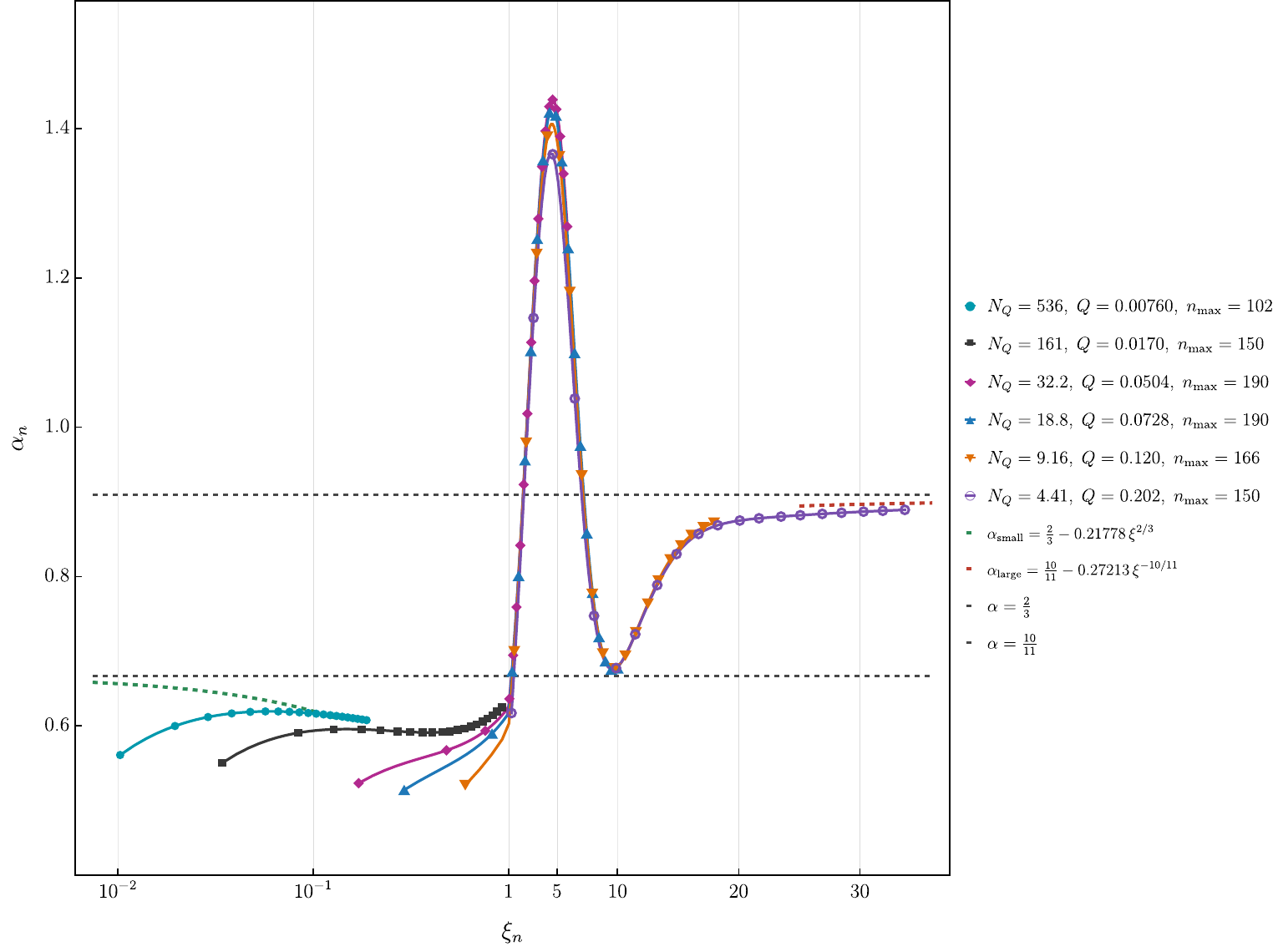}
    \caption{Numerical results for the effective exponent $\alpha_n$ as a function of the rescaled overtone number $\xi_n$ in \eqref{eq:defxin}. Dashed lines are the leading corrections \eqref{eq:leadcorrectionsalpha} to the asymptotic exponents, derived in the double-scaling limit ($n \to \infty$, $Q \to 0$). In order to exhibit both asymptotic regimes clearly, the plot switches from a logarithmic to a linear horizontal axis at $\xi_n = 1$. At intermediate values $\xi_n = \ocal(1)$, the plot exhibits a cardiogram-like structure reminiscent of \cite{Bini:2008qg}. We mark twenty equidistant modes for each curve, starting from $n=5$. Small overtones with $n = \ocal(1)$ are far from the double-scaling limit, which requires $n \to \infty$. This explains the large deviations from the limiting curve.}
    \label{fig:transition}
\end{figure}

To start with, we compute numerically the quasinormal modes of a background with an interior that transitions between two distinct Kasner epochs. The results are shown in Fig.~\ref{fig:transition},
demonstrating that the non-analytic $\frac{1}{n^\alpha}$ term in $\left.\frac{\pa \omega_n}{\pa k^2}\right|_{k=0}$ transitions, as a function of $n$, between the values of $\alpha$ corresponding to the two different Kasner epochs. In the following \S\ref{sec:analytic} we derive this crossover analytically. Finally, in \S\ref{sec:billiards} we argue that interiors with a chaotic BKL sequence of Kasner epochs  exhibit a corresponding sequence of exponents in the quasinormal modes.

The black hole background that we will use to illustrate the effect of a Kasner transition on the quasinormal modes will be a solution of a five-dimensional Einstein-Maxwell-dilaton theory,
\be\label{eq:EMD}
{\mathcal L} = \frac{1}{2\k^2} \left[R  - \frac{1}{2} g^{ab}\pa_a \phi \pa_b \phi - \frac{1}{4}Z(\phi)F^2 - V(\phi) \right]\,,
\ee
with the dilatonic coupling and potential
\be\label{eq:ZV}
Z(\phi) = e^{-\sqrt{33} \phi/3} \,, \qquad V(\phi) = - \frac{176}{25} e^{7 \phi/(2 \sqrt{33})} - \frac{132}{25} e^{-4 \phi/\sqrt{33}} + \frac{8}{25} e^{\sqrt{33} \phi/3} \,.
\ee
Here we can write $V(\phi) = \frac{1}{2} W'(\phi)^2 - \frac{1}{3} W(\phi)^2$, in terms of the superpotential
\be
W(\phi) = \frac{22}{5} e^{- 2 \phi/\sqrt{33}} + \frac{8}{5} e^{11 \phi/(2 \sqrt{33})} \,.
\ee
The potential has $V(0) = -12$, $V'(0)=0$ and $V''(0) = -4$ and therefore admits unit radius AdS$_5$ asymptotics where the scalar field has mass squared $-4$. This mass squared saturates the five-dimensional Breitenlohner-Freedman bound and corresponds to a dual operator with dimension $\Delta =2$. The potential in (\ref{eq:ZV}) is a particular case of theories considered in \cite{Arean:2024pzo}. That paper found analytic black hole solutions, which we summarise in Appendix \ref{ap:exact}. The extra ingredient in these solutions, relative to ones we have discussed so far, is that they depend on an additional parameter $Q$, which controls the charge of the black hole.

When the parameter $Q \ll 1$, so that the charge is small, the backreaction of the Maxwell field on the geometry is suppressed. Furthermore, in the solution (see Appendix \ref{ap:exact}) the scalar field is not sourced directly at the boundary, but is only turned on in the bulk due to the dilatonic coupling to the Maxwell field. It follows that when the charge is small the solution is close to the Schwarzschild-AdS$_5$ solution until deep in the interior. At sufficiently small $Q$ this means that one finds the five-dimensional Schwarzschild near-singularity Kasner exponents ($p_t = -\frac{1}{2}, p_i = \frac{1}{2}$) at intermediate interior radii. However, at the furthest depths of the interior the scalar field ultimately drives the geometry to a distinct `final' Kasner geometry as $r \to \infty$. That is, when $Q \ll 1$ the interior contains two distinct parametric Kasner regimes. The corresponding Kasner exponents are shown in Table \ref{tab:two}.
\begin{table}[h]
    \centering
    \begin{tabular}{ >{$\textstyle}c<{$}  >{$\textstyle}c<{$}  >{$\textstyle}c<{$}  >{$\textstyle}c<{$}  }
    \toprule
        \text{regime} & p_t  & p_i & \alpha \\ \midrule
       1 \ll r \ll \frac{1}{\sqrt{Q}} & -\frac{1}{2}  & \frac{1}{2}  &  \frac{2}{3} \\
  \frac{1}{\sqrt{Q}} \ll r & - \frac{7}{26} & \frac{11}{26} & \frac{10}{11}  \\ \bottomrule
    \end{tabular}
    \caption{Kasner exponents for the two different Kasner regimes present in the interior of the Einstein-Maxwell-dilaton black hole, with the horizon at $r=1$, in the limit when $Q \ll 1$.}
    \label{tab:two}
\end{table}

None of the Kasner exponents in Table \ref{tab:two} change sign across the transition. This means that this is not a canonical BKL type of transition, which always involves an interchange of collapsing and expanding dimensions. Nonetheless, it is sufficient for our purpose of demonstrating that  multiple interior Kasner epochs can be resolved in the asymptotic quasinormal modes.

To extract an effective $n$-dependent non-analytic exponent $\alpha$ from the quasinormal modes, we proceed in a similar way to our analysis in \S\ref{sec:numericsA}. There, we extracted the coefficient and here we wish to extract the exponent. We therefore take the following logarithmic derivative. Recalling that $B_n$ was defined in (\ref{eq:bn}), let
\be\label{eq:alpha}
\alpha_n \equiv
-\frac{
\log\!\left|B_{n+1}\right|
-\log\!\left|B_n\right|
}{
\log(n+1)-\log n
} \,.
\ee
This is a discrete version of $- \text{Re} \, \frac{\pa \log [\pa_n (N_n D_n)]}{\pa \log n}$.  Evaluated on the large $n$ prediction (\ref{eq:Aform}) this leads to $\alpha_n = \alpha + o(1)$, picking out the non-analytic exponent. Note that while the $1/N_n$ correction is canceled in (\ref{eq:alpha}), the $1/N_n^2$ term is not. A more complicated formula would therefore be necessary to extract any value of $\a > 2$. 

We compute the quasinormal modes of a massless scalar field in the Einstein-Maxwell-dilaton background numerically, following the method described in Appendix \ref{ap:numerical}. The logarithmic derivative (\ref{eq:alpha}) is determined from these modes and is plotted in Fig.~\ref{fig:transition}.
The most immediate result in the figure is that the effective exponent $\alpha_n$ tends towards the limiting values $\frac{2}{3}$ and $\frac{10}{11}$ expected for the two Kasner regimes in Table \ref{tab:two}. The figure shows results for several different values of small $Q$ that have been collapsed onto a single universal curve by plotting as a function of
\be
\label{eq:defxin}
\xi_n \equiv {|N_n| \over N_Q} \,,
\ee
where
\be\label{eq:NQ}
N_Q \equiv \frac{|{\mathcal Z}_+|}{\pi} \frac{(1+Q)^{4/5}}{Q^{3/2}} \,.
\ee
In five dimensions $N_n$ retains the same form (\ref{eq:modes1}), but a massless scalar now has $\nu = 2$. For $Q=0$ we have ${\mathcal Z}_+ = {\pi \over 4}(1+i)$, which is simply its value for the Schwarzschild-AdS$_5$ solution. For the values of $Q$ considered in Fig.~\ref{fig:transition}, ${\mathcal Z}_+$ changes by less than $3 \%$. We will now explain the origin of the collapse seen in  Fig.~\ref{fig:transition}, and also derive the asymptotic behaviour of the curve analytically.

\subsection{A Kasner transition: analytics}
\label{sec:analytic}

We may write down the massless wave equation in the background given in Appendix \ref{ap:exact}. The radial coordinate is such that $r \to \infty$ at the singularity, $r=1$ is the horizon and the boundary is at $r = 0$. Set $\Phi = \psi(r) \, e^{- i \omega t + i k x_1}$ in the wave equation, with no radial pre-factor here.  The small $Q$ scaling limit is obtained by taking $Q \to 0$ with the following quantities held fixed:
\be
\label{eq:scalinglimit}
\rho \equiv \frac{1}{\sqrt{Q} \, r} \,, \qquad \hat \omega \equiv Q^{3/2} \omega \,, \qquad \hat k = \sqrt{Q} \, k \,.
\ee
This limit zooms into the region of the interior, $r \sim \frac{1}{\sqrt{Q}}$ where the transition between the Kasner regimes occurs (see Table \ref{tab:two}), and the wave equation becomes
\be\label{eq:scaled}
\psi'' + \frac{\psi'}{\rho} + \left(\hat k^2 + (1 + \rho^2)^{\frac{4}{5}} \rho^{\frac{12}{5}} \hat \omega^2 \right) \psi = 0 \,.
\ee
This equation will play the role of the Kasner regime equation (\ref{eq:xeq}) that we used previously to compute the correction to the monodromy. Equation (\ref{eq:scaled}) should capture the entire universal curve in Fig.~\ref{fig:transition}. We will show explicitly that it captures the limiting behaviours.

We are looking for large $|\omega|$ quasinormal modes. However, as $Q \to 0$ these modes may have $|\hat \omega|$ large or small. We wish to write (\ref{eq:scaled}) as a Bessel equation plus a perturbation, as in (\ref{eq:xeq}).
This must be done in two different ways, depending on $|\hat \omega|$. For $|\hat \omega| \ll 1$, set $\rho  = (3 x/\hat \omega)^{1/3}$ and expand (\ref{eq:scaled}) at small $\hat \omega$ to obtain
\be\label{eq:p1}
\psi'' + \frac{\psi'}{x} + \left(1 + \frac{4}{5} \left(\frac{\hat \omega}{3 x}\right)^{\frac{2}{3}} + \cdots \right) \psi = - \frac{\hat k^2}{\hat \omega^{\frac{2}{3}}} \frac{1}{(3x)^{\frac{4}{3}}} \psi \,.
\ee
For $|\hat \omega| \gg 1$, set $\rho = (11 x/5\hat \omega)^{5/11}$ and now expand (\ref{eq:scaled}) at large $\hat \omega$ to obtain
\be\label{eq:p2}
\psi'' + \frac{\psi'}{x} + \left(1 + \frac{4}{5} \left(\frac{11 x}{5 \hat \omega}\right)^{\frac{10}{11}} + \cdots \right) \psi = - \frac{\hat k^2}{\hat \omega^{\frac{10}{11}}} \left(\frac{5}{11 x}\right)^{\frac{12}{11}} \psi \,.
\ee
If we drop the correction in the parentheses in (\ref{eq:p1}) and (\ref{eq:p2}) we recover precisely the previous equation (\ref{eq:xeq}), used to compute the correction to the near-singularity monodromy. Comparing the right hand sides of (\ref{eq:p1}) and (\ref{eq:p2}) with the right hand side of (\ref{eq:xeq}) we may read off the values $\alpha = \frac{2}{3}$ and $\alpha = \frac{10}{11}$, respectively. These agree with the values obtained from the background geometry in Table \ref{tab:two}. Thus we see that the power of the non-analytic correction to the quasinormal frequencies depends on the value of the unperturbed $\hat \omega_{0n}$, as we found numerically in Fig.~\ref{fig:transition}.

The transition between the two exponents occurs at $|\hat \w_{0n}| \approx 1$. This corresponds to $|N_n| \approx N_Q$, defined in (\ref{eq:NQ}). In the definition of $N_Q$ we include a subleading small $Q$ correction in the numerator $(1+Q)^{4/5}$. Including this correction results in a better scaling collapse at nonzero $Q$, and is natural given the form of the metric in Appendix \ref{ap:exact}. That is, we have shown that $|N_n| \gg N_Q$ leads to an $N_n^{-\frac{10}{11}}$ correction while $|N_n| \ll N_Q$ will have an $N_n^{-\frac{2}{3}}$ correction. 

Restoring the corrections in the parentheses in (\ref{eq:p1}) and (\ref{eq:p2}) we can obtain the approach to the asymptotic values of $\a$. The details of this computation are in Appendix \ref{app:analytic-alpha}. The strategy is to compute the correction to the monodromy to second order, accounting for both the perturbation due to $k^2$ (on the right hand side of the equations) and the perturbation due to the geometry (in the parentheses). In terms of the scaled overtone number $\xi_n$ in (\ref{eq:defxin}) we find
\be
\begin{aligned}
\alpha_n &= {2 \over 3} - 0.21778 \, \xi_n^{2/3} + \ldots   & \quad (\xi_n \ll 1)  \, , \\
\alpha_n &= {10 \over 11} - \frac{0.27213}{\xi_n^{10/11}}  + \ldots   &\quad (\xi_n \gg 1) \, . 
\end{aligned}\label{eq:leadcorrectionsalpha}
\ee
These expressions hold in the double-scaling limit $n \to \infty$, $Q \to 0$ with $\xi_n$ held fixed, see Appendix \ref{app:analytic-alpha}. In this limit the transition is populated by a large number of modes. In Fig.~\ref{fig:transition} the analytical results (\ref{eq:leadcorrectionsalpha}) are seen to be in good agreement with the numerical modes.

\subsection{Signatures of cosmological billiards}
\label{sec:billiards}

Fig.~\ref{fig:transition} suggests that the many distinct Kasner epochs of an interior undergoing BKL chaos will be audible in the asymptotic quasinormal modes. In this section we obtain the boundary signature of a simplified interior BKL dynamics. As we noted in the introduction, the full BKL scenario involves highly inhomogeneous spacetimes that are beyond the technology we have developed in this paper \cite{Belinsky:1970ew, Damour:2002et, Garfinkle:2003bb, Garfinkle:2020lhb}. The mixmaster universe is a homogeneous model \cite{Misner:1969hg} that captures essentially the same chaotic sequence of Kasner epochs as BKL (the underlying hyperbolic billiard domain is larger \cite{Belinski:2017fas}). Mixmaster chaos was engineered in a planar asymptotically AdS black hole in \cite{DeClerck:2023fax}, and the results in this section apply directly to that background.

The discussion below continues to rely on the assumptions made in \S\ref{sec:qnm}, now in a more nontrivial setting. Namely, we assume that the anti-Stokes lines connect the boundary, singularity and horizon in a certain way and that we can make a contour deformation from the appropriate anti-Stokes line to the real axis. An even more fundamental assumption is that ${\mathcal Z}_+$ is finite.\footnote{The optical depth diverges if the singularity becomes null. This can happen if the Kasner exponent $p_t \to 1$, although the limit is not always continuous \cite{Hartnoll:2020rwq}. Values of $p_t$ parametrically close to one, potentially arising on intermediate Kasner epochs, could also cause difficulties for our analysis as $\alpha$ in (\ref{eq:a12}) becomes very large.} With these assumptions in hand, we can
recap a few facts about BKL dynamics.

At late interior times, in a BKL regime, the evolution of the metric components can be mapped onto a hyperbolic billiard dynamics \cite{Damour:2002et}. The Kasner epochs describe the flight of the billiard ball from one boundary wall to the next. The bounces off the walls become increasingly sharp towards the singularity and take a negligible proper time compared to the time of flight. In the billiard regime the metric is well-approximated by a sequence of Kasner epochs with changing exponents, letting $\vec p \equiv \{p_t,p_i\}$,
\be\label{eq:sequence}
\cdots \to \;\;   \tau_{(m-1)}  \;\; \xrightarrow{\vec p_{(m-1)}} \;\; \tau_{(m)} \;\;  \xrightarrow{\vec p_{(m)}} \;\;  \tau_{(m+1)}   \;\;  \xrightarrow{\vec p_{(m+1)}} \;\; \tau_{(m+2)} \;\; \to  \cdots  \,,
\ee
where $\tau_{(m)}$ is the proper time to the singularity of the $m$th Kasner transition and $\vec p_{(m)}$ are the Kasner exponents of the $m$th Kasner epoch.

How long is the $m$th Kasner epoch? While the exact timing of the transitions is chaotic, this question can be answered statistically. Firstly note that whenever the Kasner exponents obey our assumption (\ref{eq:sum}) then, from the metric (\ref{eq:genK}), the local volume element of a spatial slice obeys $\sqrt{\gamma} \propto \tau$. It is well-known that the volume element decays doubly exponentially during BKL dynamics, see \cite{DeClerck:2023fax} for a review.
A useful version of this statement was shown in \cite{Blacker:2026poc}: Let $T$ be the proper time of hyperbolic billiard dynamics (this is not the same as $\tau$ above). To leading order as $T \to \infty$, the volume $\sqrt{\gamma} \to 0$ as
\be
\log \log \frac{1}{\sqrt{\gamma}} = 2 T \,.
\ee
Because the billiard domain has order one size, na\"ively the $m$th collision occurs on average at $T_{(m)} \sim m$. In fact there is a logarithmic correction due to long incursions into the cusp of the hyperbolic domain and, for the original four-dimensional mixmaster, $T_{(m)} \sim \frac{m}{\log m}$.\footnote{A heuristic argument goes as follows. The spatial volume decays doubly exponentially over a Kasner `era,' where each era has $k$ Kasner epochs, with $k$ drawn from a distribution with a tail falling off like $P(k) \sim 1/k^2$ \cite{LIFSHITZ1992659}. The typical number of epochs after $N$ eras is then of order $N \log N$. To obtain this finite answer from the heavy-tailed distribution we introduced a cutoff $k_\text{max} \sim N$, the maximum number of epochs expected in a sample of $N$ eras. To get to $m$ epochs one therefore only needs $\frac{m}{\log m}$ eras.} It follows that the $m$th transition occurs, typically and to leading order, at
\be\label{eq:expexp}
\log \log \frac{1}{\tau_{(m)}} \sim \frac{m}{\log m} \,.
\ee
In particular, this means that generically
\be\label{eq:hierarchy}
\tau_{(m+1)} \ll \tau_{(m)} \,.
\ee

Within each Kasner epoch, the proper time and the Schr\"odinger coordinate $x$ are related by a power-law (\ref{eq:xx}). It follows that the doubly-exponential hierarchy (\ref{eq:hierarchy}) remains true for $x$:
\be\label{eq:hier2}
x_{(m+1)} \ll x_{(m)} \,.
\ee
Now write the Schr\"odinger equation (\ref{eq:schr}) as
\be\label{eq:se5}
- \frac{d^2 \Psi}{dx^2} = \left(\omega^2 - V(x)\right) \Psi \,.
\ee
Here the Schr\"odinger coordinate $x \equiv \int \frac{dr}{h(r)}$, generalising (\ref{eq:xx}) away from any particular Kasner regime. For $x_{(m+1)} \ll x \ll x_{(m)}$ the geometry will be Kasner and hence we know that the potential takes the form $V_K(x)$ in (\ref{eq:V3}), with an exponent $\alpha_{(m)}$ corresponding to the $m$th Kasner epoch. Define $\omega_{(m)} \equiv \frac{1}{x_{(m)}}$ and suppose that we wish to obtain the correction to the monodromy for a quasinormal frequency $\omega = \omega_{0n}$ with
\be\label{eq:wrange}
\omega_{(m)} \ll |\omega_{0n}| \ll \omega_{(m+1)} \,.
\ee
Now set $x = \bar x/\omega_{0n}$ and let $\Psi = \sqrt{x} \psi$. In the range $|\omega_{0n}| x_{(m+1)} \ll \bar x \ll |\omega_{0n}| x_{(m)}$ the Schr\"odinger equation becomes
\be
\psi'' + \frac{\psi'}{\bar x} + \psi = - \frac{\vep}{(\omega_{0n})^{\a_{(m)}}} \bar x^{\a_{(m)}-2} \psi \,,
\ee
where $\vep$ was defined below (\ref{eq:V3}). This equation is recognised as the familiar (\ref{eq:xeq}). Crucially, because the frequency is in the range (\ref{eq:wrange}) we have $|\omega_{0n}| x_{(m+1)} \ll 1$ and $1 \ll |\omega_{0n}| x_{(m)}$. This means that $\bar x$ can be much less or much greater than one, giving it the dynamical range needed for the monodromy computation of \S\ref{sec:qnm2}.

It follows from the above discussion that the exponent of the non-analytic term in $D_n$ tracks the corresponding Kasner epoch of the BKL dynamics. Transitions occur whenever $\omega_{0n}$ exits a range of the form (\ref{eq:wrange}). That is to say, transitions occur at
\be\label{eq:Nx}
n_{(m)} \sim \frac{|{\mathcal Z}_+|}{\pi x_{(m)}} \,.
\ee
This equation is a generalisation of (\ref{eq:NQ}). Recall from Fig.~\ref{fig:transition} that the transition does not occur at a single mode, but across many modes. For this reason (\ref{eq:Nx}) is lacking a precise numerical prefactor. Nonetheless, Kasner transitions can be described analytically \cite{Belinsky:1970ew, DeClerck:2023fax} and are increasingly abrupt at late interior times. By zooming in on a Kasner transition, similarly to what was done in \S\ref{sec:analytic}, it will likely be possible to count the modes associated to the transition and thereby upgrade (\ref{eq:Nx}) to an equality.

The use of ${\mathcal Z}_+$ in (\ref{eq:Nx}) also requires a qualification. We have just seen that in order to pick out the $m$th Kasner epoch in the quasinormal spectrum one must
compute the monodromy with $|x|$ in the range $x_{(m+1)} \ll |x| \ll x_{(m)}$. That is, the WKB contour $A$ in Fig.~\ref{fig:rplane} does not reach all the way to the vicinity of the singularity at $x=0$. The truncated contour affects the leading order mode computation in \S\ref{sec:largen}, producing an $m$-dependent `partial' optical distance, ${\mathcal Z}_{+(m)}$, which need not coincide with the full ${\mathcal Z}_+$.
However, the optical depth between successive Kasner transitions is $x_{(m)}-x_{(m+1)}
\approx x_{(m)} \ll 1$. Late-interior epochs therefore make only a small contribution to the full optical distance, so that ${\mathcal Z}_{+} \approx {\mathcal Z}_{+(m)}$ up to exponentially small corrections. Note, however, that while the epoch-dependent shifts in the quasinormal mode spacing are small, the shift in the phase $\omega ({\mathcal Z}_{+} - {\mathcal Z}_{+(m)})$ need not be.

Recall that $x_{(m)}$ is vanishing towards the singularity and hence $n_{(m)}$ is diverging. Therefore, corresponding to the sequence (\ref{eq:sequence}) of Kasner exponents we have the sequence
\be
\cdots \to \;\;   n_{(m-1)}  \;\; \xrightarrow{N_n^{-\a_{(m-1)}}} \;\; n_{(m)} \;\;  \xrightarrow{N_n^{-\a_{(m)}}} \;\;  n_{(m+1)}   \;\;  \xrightarrow{N_n^{-\a_{(m+1)}}} \;\; n_{(m+2)} \;\; \to  \cdots  \,.
\ee
More explicitly, $D_n = d(\a_{(m)}) N_n^{-\a_{(m)}} + \cdots$ for $n_{(m)} \ll n \ll n_{(m+1)}$. The entire chaotic sequence of BKL transitions is therefore audible within the quasinormal mode spectrum.

From (\ref{eq:expexp}), (\ref{eq:xx}) and (\ref{eq:Nx}) we have that
\be\label{eq:expexp2}
\log \log n_{(m)} \sim \frac{m}{\log m} \,.
\ee
That is, doubly exponentially large mode numbers are needed to probe increasingly late Kasner epochs. Equations (\ref{eq:Nx}) and (\ref{eq:expexp2}) are quantitative instances of the UV/UV correspondence discovered in \cite{Festuccia:2005pi}. Highly excited quasinormal modes are needed to probe the short-distance near-singularity metric in the bulk. Because the Kasner transitions accumulate doubly exponentially close to the singularity, a doubly exponentially large mode number is needed to resolve them.

\section{Discussion}
\label{sec:discuss}

In this paper we have shown that intrinsic geometric data of the black hole singularity is well-captured by the dispersion of the quasinormal modes $\left. \frac{\pa \omega_n}{\pa k^2} \right|_{k=0}$ at large overtone number $n \gg 1$. In particular, we have identified quantitative signatures \eqref{eq:intro} and \eqref{eq:introd} of Kasner epochs in the interior and furthermore shown in Fig.~\ref{fig:transition} that transitions between Kasner epochs are audible as a function of the overtone number $n$. 
We have studied nonzero-temperature black hole geometries, whose quasinormal mode frequencies can be interpreted as the eigenvalues of a modified boost generator \cite{Chen:2023hra}. Our simple results, then, may suggest
the following general picture: the near-singularity geometry encodes the properties of the high-energy spectrum of the modified boost operator in strongly coupled large-$N$ quantum systems. 

There are some conceptually straightforward generalisations of the computations we have performed. One can look at the large overtone quasinormal modes for metric fluctuations and other fields, and in black holes with spherical rather than planar horizons.
One can also explore the interiors of charged and rotating black holes, in the spirit of e.g.~\cite{AliAhmad:2026wem,Ceplak:2025dds}. Note, however, that deformations of the theory by scalar operators typically remove the Cauchy horizons of charged black holes, at least, and render these interiors similar to the neutral cases we have discussed here \cite{Hartnoll:2020rwq}. Moving beyond AdS/CFT, it would be very interesting
to understand how the interior geometry of astrophysical, asymptotically flat (or de Sitter) black holes is encoded in the asymptotic quasinormal modes. See e.g.~\cite{Grozdanov:2026ktq, Grozdanov:2026lnc} for recent related discussions. A non-analytic correction to the quasinormal modes of an asymptotically flat, four-dimensional black hole was obtained in \cite{MaassenvandenBrink:2003as}. The 
derivative ${\pa \omega_n \over \pa \lambda_{\ell}} \Big|_{\ell=0}$ with respect to the quadratic Casimir $\lambda_{\ell} = \ell (\ell + 1)$ of the modes in (A.6) of that paper recovers our exponent $\alpha$.

It remains to flesh out the discussion of BKL signatures in \S\ref{sec:billiards}. In the first instance this could be a numerical check of the presence of multiple transitions in the quasinormal mode spectrum, using an explicit homogeneous model such as \cite{DeClerck:2023fax}. Further afield, the sequence of Kasner exponents in a cosmological billiards trajectory is constrained by powerful modular symmetries. In four dimensional pure gravity the billiard is half the fundamental domain of $SL(2,\Z)$ \cite{Belinski:2017fas} and sequential bounces are controlled by the Gauss map \cite{PhysRevLett.46.963, Khalatnikov1985}. Remarkably, 
the cosmological billiards of ten and eleven dimensional supergravity are also subject to exotic modular symmetries, see e.g.~\cite{PhysRevLett.85.920, PhysRevLett.86.4749, Feingold:2008ih, Kleinschmidt:2010bk}. It would be fascinating to understand how these symmetries are manifested in the quasinormal modes, and whether this gives a clue to their dual field theoretic meaning.

In terms of finding boundary signatures of generic BKL dynamics, a further big open direction is to allow inhomogeneous sources in the boundary theory. A first step in that direction could be to include boundary spatial inhomogeneity but to retain boundary time-translation symmetry. The language of quasinormal modes will then carry over unchanged, except that modes with different momenta will couple. That is, the Green's function will have the form $G_{12}(\omega,\vec x_1, \vec x_2)$.
A speculative possibility is that the BKL decoupling of spatial points $\vec x$ close to the singularity means that the large overtone quasinormal modes have non-analytic corrections of the form $n^{-\a(\vec x)}$. In the Green's function, $\vec x(\vec x_1, \vec x_2)$ could be the bulk point closest to the singularity on the bouncing geodesic connecting $\vec x_1$ and $\vec x_2$.

It should be noted that the BKL scenario is not the only possibility for generic near-singularity dynamics. Indeed, there is an extensive body of mathematical work on null singularities, especially in the context of charged and rotating black holes. Some older and newer results can be found in \cite{PhysRevD.41.1796, Luk2017,10.4007/annals.2025.202.2.1, moortel}. The dual holographic implications of this literature remain unexplored.

Quasinormal modes are known to be unstable against small perturbations of the effective Schr\"odinger potential \cite{Cheung:2021bol,Arean:2023ejh}: perturbing the potential slightly can create new quasinormal modes and shift the existing modes by a relatively large amount. Given that the quasinormal mode dispersion studied in the paper is controlled by the local geometry of the singularity, we expect that such perturbations will not affect the basic conclusions of our paper. However, such instabilities could create additional subtleties when applying our results at finite $n$ and we leave their better understanding for future work.

Finally, the most compelling challenge arising from our results is to recover the signatures of the singularity from a boundary field theory computation, and then to understand their fate at finite 't Hooft coupling and/or $N$. The connection, mentioned above, to the modified boost operator \cite{Chen:2023hra} may allow these questions to be framed in terms of the algebraic properties of quantum systems, in the spirit of \cite{Leutheusser:2021frk,Witten:2021jzq,Witten:2021unn,Leutheusser:2025zvp}.

The analytic structure of thermal correlation functions at finite $N$ and 't Hooft coupling, and hence the fate of quasinormal modes, remains poorly understood \cite{Hartnoll:2005ju}. At finite 't Hooft coupling, one challenge could be the proliferation of different branches of quasinormal modes, as is seen in an SYK analysis \cite{Dodelson:2024atp}.
Moving to finite $N$ may be cleaner
in settings where the boundary theory has infinite spatial volume, so that there is still a thermodynamic limit. For example, it
may be possible to understand the fate of quasinormal modes in tractable non-holographic large $N$ models such as SYK chains \cite{Choi:2020tdj,Dodelson:2026gak}. 
The simplest holographic model exhibiting quasinormal modes may be BFSS matrix quantum mechanics \cite{Biggs:2023sqw}. This $0+1$ dimensional theory is not in an infinite volume limit, however, and so the finite $N$ fate of the modes is more subtle. At finite $N$, the local volume element of the interior will reach the Planck scale at some $\left. x\right|_\text{Pl}$ which, very schematically from (\ref{eq:Nx}), will correspond to a mode number $\left. n\right|_\text{Pl} \sim N$, possibly to some power. What happens beyond that point? Does quantum gravity make the music stop, or does a new Planckian music ring out?

\section*{Acknowledgements}

This work has been partially supported by STFC consolidated grant ST/T000694/1. SAH is partially supported by Simons Investigator award \#620869. The numerical code used to compute quasinormal modes has been written by OpenAI’s ChatGPT and Codex agents. The analytical calculations and writing of the paper have been human-generated, with assistance from the agents.

\appendix

\section{Corrections to the near-singularity scaling}
\label{ap:cor}

The equations of motion for the scalar field $\phi$ and metric components $f,\chi$ following from the Einstein-scalar action (\ref{eq:ES}) are \cite{Frenkel:2020ysx}
\begin{align}
\chi' & = \frac{r}{2} \phi^{\prime 2} \,, \label{eq:chieq} \\
4 r f' & = 12 (f-1) - 2 \phi^2 + r^2 f   \phi^{\prime 2} \,, \\
\phi'' & = \left(\frac{2}{r} + \frac{\chi'}{2} - \frac{f'}{f}\right)\phi' - \frac{2}{r^2 f} \phi \,.
\end{align}
The near-Kasner expansion then takes the form
\begin{align}
f & \textstyle = - f_o r^{4/(1 - p_t)} + \frac{3 (1+p_t)(1 + 3 p_t)}{4}\log^2r + \frac{(1 - p_t)(1 + 3 p_t)^2}{4} \log r  - \frac{(1-p_t)(27 p_t^3 - 15 p_t^2 - 47 p_t - 37)}{32} + \cdots \,, \nonumber \\
\phi & \textstyle = 2 \sqrt{\frac{1 + 3p_t}{1 - p_t}}\log r + \frac{\sqrt{(1-p_t)(1 + 3 p_t)}}{r^{4/(1 - p_t)}} \left(-\frac{1 + 3 p_t}{4 f_o} \log^2r - \frac{3 p_t (1 - p_t)}{4 f_o} \log r + \frac{(1 - p_t)(9 p_t^2-10 p_t-11)}{32 f_o} \right) + \cdots\,, \nonumber \\
\chi & \textstyle = 2 \frac{1+ 3p_t}{1 - p_t} \log r + \chi_o+\frac{1 + 3 p_t}{r^{4/(1 - p_t)}} \left(
- \frac{1 + 3 p_t}{2 f_o}\log^2 r - \frac{3p_t(1 - p_t)}{2 f_o} \log r + \frac{(1 - p_t)(9 p_t^2 - 10 p_t -11)}{16 f_o} \right) + \cdots \,. \nonumber
\end{align}
While the logarithmic corrections vanish in the Schwarzschild limit $p_t = -\frac{1}{3}$, the constant correction to $f$ remains nonzero.

From the above expansions, the quantities (\ref{eq:V}) appearing in the wave equation are
\begin{align}
V & \textstyle = - h_o^2 r^4 \Big(1 + \frac{k^2}{f_o r^{2(1+p_t)/(1-p_t)}} + \frac{1}{2 f_o r^{4/(1 - p_t)}}  \nonumber \\
& \textstyle \quad  \times \left[ 
\frac{2(1+p_t)(1 + 3 p_t)}{1 - p_t} \log^2 r -
(1-p_t)(1 + 3 p_t) \log r + \frac{8 m^2 - 3 p_t^3 + 5 p_t^2 + 11 p_t + 11}{4}
\right]\Big) + \cdots\,, \nonumber \\
h & \textstyle = - h_o r^3\left( 1 - \frac{1- p_t}{4 f_o r^{4/(1 - p_t)}} \left[\frac{2(1 + 3 p_t)}{1 - p_t} \log^2 r +
(1 + 3 p_t) \log r - \frac{3 p_t^2 - 2 p_t-13}{4} \right]\right) + \cdots \,. \nonumber
\end{align}
The corrections here define the non-universal functions $F_V$ and $F_h$ appearing in (\ref{eq:VK0}). 
Again, the logarithmic corrections vanish in the Schwarzschild limit but the constant corrections remain. In particular, putting $p_t = - \frac{1}{3}$ gives the Schwarzschild results
\begin{align}
V_\text{Schw} & \textstyle = - h_o^2 r^4 \Big(1 + \frac{k^2}{f_o r} + \frac{1 + m^2}{f_o r^{3}} \Big) + \cdots\,, \nonumber \\
h_\text{Schw} & \textstyle = - h_o r^3 \left(1 - \frac{1}{f_o r^3}\right) + \cdots \,. \nonumber
\end{align}
In terms of the Schr\"odinger coordinate $x = \int \frac{dr}{h}$, one has $r = \frac{1}{\sqrt{2 h_o x}} + \frac{2 h_o x}{5 f_o} + \cdots$ and the potential
\be
V_\text{Schw} = -\frac{1}{4 x^2} - \frac{\sqrt{h_o}}{2 \sqrt{2} f_o} \frac{k^2}{x^{3/2}} - \frac{h_o^{3/2}}{\sqrt{2} f_o} \frac{\frac{9}{5} + m^2}{x^{1/2}} + \cdots \,.
\ee
The coefficient of the $m^2$ term here only depends on the leading order near-singularity Kasner geometry, but
the shift of the mass squared by $\frac{9}{5}$ in the potential is due to the non-universal corrections obtained above. This fact was commented on in the main text below (\ref{eq:wfull}). 

\section{Two-sided correlator via the product formula}
\label{ap:two}

In this appendix we obtain the leading non-analytic correction to the large $E$ two-sided correlator (\ref{eq:g12}), using the thermal product formula of \cite{Dodelson:2023vrw} for $G_{12}(\omega)$, see also \cite{Horowitz:2023ury,Grozdanov:2024wgo,
Bhattacharya:2025vyi,Grozdanov:2025ulc}. In \cite{Dodelson:2023vrw} it was already shown how the Green's function at large real $\omega$, including non-analytic corrections of the kind obtained in \cite{Afkhami-Jeddi:2025wra}, could be computed from the asymptotic quasinormal modes. The 
computation below is very similar, now with an emphasis on large imaginary $\omega = - i E$.

Let us write the large overtone quasinormal modes (\ref{eq:wfull}) as
\be\label{eq:wn}
{\mathcal Z}_+ \omega_n = \pi n + a + b \left(\frac{{\mathcal Z}_+}{\pi n}\right)^\alpha + \cdots \,,
\ee
with complex coefficients $a,b$ and real exponent $\alpha$. The thermal product formula implies that
\be\label{eq:qnmsum}
\pa_\omega \log G_{12}(\omega) = - \sum_{n=1}^\infty \frac{2 \, \omega}{\omega^2 - \omega_n^2} + \omega_n \leftrightarrow \omega_n^* \,,
\ee
and setting $\omega = - i E$,
\be\label{eq:sum2}
\pa_E \log G_{12}(-i E) = - 2 \, \text{Re} \sum_{n=1}^\infty \frac{2 E}{E^2 + \omega_n^2}\,.
\ee
We will now extract the leading analytic and non-analytic terms from the sum (\ref{eq:sum2}).

Consider some
lower cutoff $\Lambda$, independent of $E$ and such that the asymptotic form (\ref{eq:wn}) holds to good approximation for $n > \Lambda$. At large $E$ we can write
\begin{align}
\sum_{n=1}^\infty \frac{2 E}{E^2 + \omega_n^2} & =  \frac{2}{E} \sum_{n=\Lambda}^\infty \frac{1}{1 + {\mathcal Z}_+^{-2}\left(\frac{\pi n}{E} + \frac{a}{E} + \frac{b}{E^{1+\a}} \left(\frac{\pi n}{{\mathcal Z}_+ E} \right)^{-\alpha}\right)^2} + \ocal\left(\frac{1}{E}\right) \\
& = \frac{2}{E}  \sum_{n=\Lambda}^\infty \frac{1}{1 +  \left( \frac{\pi n}{{\mathcal Z}_+ E}\right)^2} - \frac{4 b}{{\mathcal Z}_+ E^{2+\alpha}} \sum_{n=\Lambda}^\infty \frac{\left(\frac{\pi n}{{\mathcal Z}_+ E}\right)^{1-\alpha}}{\left(1 + \left( \frac{\pi n}{{\mathcal Z}_+E}\right)^2\right)^2} + \ocal\left(\frac{1}{E}\right) \,.
\end{align}
The $a$ term has been incorporated into the analytic part of the expansion.
At large $E$ the sum can be approximated by an integral, so that
\begin{align}
\sum_{n=1}^\infty \frac{2 E}{E^2 + \omega_n^2}  & = \frac{2 {\mathcal Z}_+}{\pi} \int_0^\infty \frac{dx}{1 + x^2} - \frac{4 b}{\pi E^{1+\alpha}} \int_0^\infty \frac{x^{1-\alpha} dx}{\left(1 + x^2\right)^2} +\ocal\left(\frac{1}{E}\right) \,, \\
& = {\mathcal Z}_+ -  \a \csc \frac{\pi \a}{2} \frac{b}{E^{1+\a}}   + \ocal\left(\frac{1}{E}\right) \,. \label{eq:qnmsum2}
\end{align}
In the integrals we changed variables to $n = \frac{{\mathcal Z}_+ E}{\pi} x$. Note that $\Lambda/E \to 0$ and hence the lower limit of integration can be taken to zero so long as the integrals converge (which they do for $-2 < \a < 2$, and we analytically continue otherwise).

Using (\ref{eq:qnmsum2}) in (\ref{eq:sum2}), and taking the value of $b$ from (\ref{eq:wfull}), we have that
\be
\pa_E \log G_{12}(- i E) = - 2\, \text{Re} \,{\mathcal Z}_+  + \frac{\vep}{8} \frac{\pi^{3/2} \Gamma\left(\frac{1-\a}{2}\right)}{\Gamma\left(1 - \frac{\a}{2} \right)^3} \frac{\a (2 h_o)^\a}{E^{1+\a}} + \cdots \,.
\ee
Integrating this expression up, and keeping track only of the non-analytic correction to the leading exponential result,
\be\label{eq:Ecorrect}
G_{12}(-i E) = e^{-2 E \,\text{Re}\,{\mathcal Z}_+}\left(1 - \frac{\vep}{8} \frac{\pi^{3/2} \Gamma\left(\frac{1-\a}{2}\right)}{\Gamma\left(1 - \frac{\a}{2} \right)^3} \frac{(2 h_o)^\a}{E^{\a}} + \cdots \right) \,.
\ee
Thus we have re-obtained (\ref{eq:g12}), together with the leading non-analytic correction.

The leading order correspondence between the geodesic result $G_{12} \approx e^{-2 E \,\text{Re}\,{\mathcal Z}_+}$ and the asymptotic frequencies $\omega_n \approx \frac{\pi n}{{\mathcal Z}_+}$ implies that any failure of the anti-Stokes line assumptions in \S\ref{sec:largen} must also manifest as a failure of the na\"ive geodesic saddle point. Related to this observation, the correction (\ref{eq:Ecorrect}) can also be obtained directly from a WKB computation of the large $E$ correlator. However, WKB perturbation theory breaks down when the turning point is deep in the near-singularity Kasner regime and one is forced to do an anti-Stokes line analysis very similar to the quasinormal mode computation of \S\ref{sec:largen} and \S\ref{sec:qnm2}. A conventional second order WKB computation (as was done for the large mass limit in \cite{Frenkel:2020ysx}) will get the correct non-analytic power but not the correct prefactor, which involves four gamma functions in (\ref{eq:Ecorrect}).

\section{Details of exact backgrounds}
\label{ap:exact}

\subsection*{Einstein-scalar theory with potential}

The two exact background solutions of the theory defined by the action (\ref{eq:ESV}) are as follows. In both cases the geometries are written most conveniently in the form
\be\label{eq:einsv}
ds^2 = \Omega(r) \left(- g(r) dt^2 + \frac{dr^2}{g(r)} +dx_1^2 + dx_2^2 \right) \,.
\ee
This is slightly different from the parametrisation in (\ref{eq:back}). We will continue to use $r$ for the radial coordinate although, as we note shortly, its range will be different. The scalar field is $\phi = \phi(r)$. We consider two cases, with different values of $\lambda$ in the potential:

\paragraph*{Case with $\lambda = -\frac{20}{37}$:} There is a solution to the equations of motion with
\be
\Omega = \frac{9 r^2}{(r^3-1)^2} \,, \quad \phi = 2 \sqrt{2} \log r \,, \quad g = \frac{(2-r)(5 + 12 r + 6 r^2 + 8 r^3 + 4 r^4 + 2 r^5)}{37 r} \,.
\ee
The boundary is at $r=1$, the horizon is at $r=2$ and the singularity is at $r \to \infty$. The metric tends to AdS$_4$ near the boundary while $\phi \to 0$ with the required falloff. Near the singularity,
\be
ds^2 \approx \frac{18 r}{37} dt^2 - \frac{333}{2 r^9} dr^2 + \frac{9}{r^4}\left(dx_1^2 + dx_2^2 \right) \,.
\ee
This metric leads to the parameters quoted in the row \texttt{EinSV}$_1$ in Table \ref{tab:models}.

\paragraph*{Case with $\lambda = \frac{320}{71}$:} There is now a solution with
\be
\Omega = \frac{9 r^2}{(r^3-1)^2}  \,, \quad \phi = 2 \sqrt{2} \log r \,, \quad g = \frac{(2 r - 1)(16 r^5 + 8 r^4 + 4 r^3 - 78 r^2 - 39 r + 160)}{71 r}  \,.
\ee
Note that the conformal factor and scalar field are the same. The boundary is again at $r=1$ but now the horizon is at $r= \frac{1}{2}$ and the singularity is at $r \to 0$. The solution is again asymptotically AdS$_4$. Near the singularity, letting $\bar r \equiv \frac{1}{r} \to \infty$,
\be
ds^2 \approx \frac{1440}{71 \bar r} dt^2 - \frac{639}{160 \bar r^7} d\bar r^2 + \frac{9}{\bar r^2} \left(dx_1^2 + dx_2^2 \right) \,.
\ee
This metric leads to the parameters quoted in the row \texttt{EinSV}$_2$ in Table \ref{tab:models}.

\subsection*{Einstein-Maxwell-dilaton theory}

The exact background of the theory (\ref{eq:EMD}) has the metric 
\be
 ds^2=\frac{H(r)^{4/15}}{r^2}
 \left[-\frac{G(r) dt^2}{H(r)^{4/5}}
 +\frac{dr^2}{G(r)}+dx_1^2 + dx_2^2 + dx_3^2\right] \,.
 \label{eq:trmetric}
\ee
Here the radial functions are
\begin{align}
 H(r) =1+Q r^2 \,, \qquad  G(r) = H(r)^{4/5}-(1+Q)^{4/5} r^4 \,.
 \label{eq:transition-HKG}
\end{align}
The AdS$_5$ boundary is at $r=0$, the horizon is at $r=1$ and the singularity is at $r \to \infty$. The dilaton and Maxwell potential are
\be
e^{-\phi} = H(r)^{2\sqrt{33}/15} \,, \qquad A_t = \frac{2\sqrt{5 Q}}{5(1+Q)^{3/5}} \frac{1 - r^2}{1 + Q r^2} \,.
\ee

In this solution, $Q$ is a free parameter that controls the charge of the black hole. When $Q \ll 1$ there are two distinct Kasner regimes, with metrics
\be
ds^2 \approx r^2 dt^2 - \frac{dr^2}{r^6} + \frac{dx_1^2 + dx_2^2 + dx_3^2}{r^2}  \qquad (\text{for}\;\; 1 \ll r \ll \textstyle \frac{1}{\sqrt{Q}}) \,,
\ee
this is just the Schwarzschild-AdS$_5$ near-singularity Kasner geometry, and
\be
ds^2 \approx \frac{r^{14/15} dt^2}{Q^{8/15}} - \frac{Q^{4/15} dr^2}{r^{82/15}} + \frac{Q^{4/15}}{r^{22/15}}\left(dx_1^2 + dx_2^2 + dx_3^2 \right) \qquad (\text{for}\;\; \textstyle \frac{1}{\sqrt{Q}} \ll r) \,.
\ee
These two Kasner metrics lead to the parameters quoted in Table \ref{tab:two}.

\section{Numerical methods}
\label{ap:numerical}

We will numerically compute the quasinormal modes using a matched asymptotic expansion method. Many variants of this method have been used in the past for computing quasinormal modes of AdS black holes, going back to the pioneering paper \cite{Horowitz:1999jd}. There are small differences for the various backgrounds we consider, so we describe them separately.

\subsection*{Schwarzschild-AdS$_4$}

In Eddington-Finkelstein coordinates the metric is
\be
ds^2 = \frac{1}{r^2} \left[- (1-r^3) dv^2 - 2 dv dr + dx_1^2 + dx_2^2\right] \,.
\ee
We wish to find solutions to the wave equation that are infalling at the horizon and normalisable at the AdS$_4$ boundary. To this end we write the massless probe scalar field as
\begin{equation}
 \Phi(v,r,x_1)=e^{-i \omega v+i k x_1}\,r^3 y(r) \,.
 \label{eq:scalar-ansatz}
\end{equation}
The wave equation (\ref{eq:wave}), with $m^2=0$, then becomes
\be
 r(1-r^3)y''+\bigl(4-7r^3+2 i \omega r\bigr)y'
 +\bigl(4 i \omega-qr-9r^2\bigr)y=0 \,.
 \label{eq:sch-ode}
\ee
Here we set $q \equiv k^2$, for convenience.

In (\ref{eq:scalar-ansatz}) we have removed the leading singular behaviour near the horizon and near the boundary.
We may therefore consider the following two approximate solutions to (\ref{eq:sch-ode}), defined by their regular series expansions near the boundary and  near the horizon, respectively, 
\be\label{eq:series}
y_B(r) = \sum_{j=0}^{j_\text{max}} b_j r^j \,, \qquad y_H(r)=\sum_{j=0}^{j_\text{max}}c_j (1-r)^j \,.
\ee
Here we can set $b_0 = c_0 = 1$ and all of the remaining coefficients are fixed recursively by the differential equation (\ref{eq:sch-ode}). The coefficients, and hence $y_B(r)$ and $y_H(r)$, are functions of $\omega$ and $q$. The remaining singular points of the equation (\ref{eq:sch-ode}) are at $r = e^{\pm i \frac{ 2\pi}{3}}$. This means that the series expansions (\ref{eq:series}) both converge at the midpoint $r = \frac{1}{2}$. We may therefore use the expansions to compute the Wronskian at the midpoint
\be\label{eq:W}
W(\omega,q) \equiv y_B'(\half)y_H(\half) - y_B(\half)y_H'(\half) \,.
\ee
At the quasinormal frequencies the solution is regular at both the boundary and the horizon, and hence the Wronskian vanishes,
\be
W(\omega_n(q),q) = 0 \,.
\ee

We firstly compute the quasinormal frequencies at $q=0$, using the Newton method to find the zeros of the Wronskian. The seed value is the WKB prediction $\omega^{(0)}_n = \omega^{\text{WKB}}_n(0)$, computed in the main text. We then iterate using the usual formula
\begin{equation}
 \omega_n^{(s+1)}=\omega_n^{(s)}-
 \frac{W(\omega_n^{(s)},0)}{\partial_\omega W(\omega_n^{(s)},0)} \,.
 \label{eq:newton}
\end{equation}
In our calculation we used $j_{\text{max}} = 240 + 10 n$ as the cutoff on the sums (\ref{eq:series}) and a Newton tolerance of $10^{-70}$. The error on the quasinormal modes was estimated by evaluating the Wronskian at different $r \neq \frac{1}{2}$, as well as by changing the cutoff. 

Once the $q=0$ quasinormal mode $\omega_n(0)$ has been found we may compute the derivative $\omega_n'(0)$ as follows. From the fact that $W(\omega(q),q) = 0$ for quasinormal modes we have
\be
{d \over d q} W(\omega,q) = {d \omega \over d q} \partial_{\omega} W(\omega,q) + \partial_{q} W(\omega,q) = 0 \,,
\ee
and hence
\begin{equation}
 D_n=\left.\frac{d\omega_n}{d q}\right|_{q=0}
 =-\left.\frac{\partial_{q} W}{\partial_{\omega} W}\right|_{(\omega_n(0),0)} \,.
 \label{eq:tangent}
\end{equation}
The final derivatives here can be calculated explicitly from (\ref{eq:W}), and then evaluated at $\omega = \omega_n(0)$ and $q=0$. In this way we obtain the $D_n$ used to evaluate (\ref{eq:Delta}) in the main text.

\subsection*{Exact backgrounds with $\alpha = \frac{3}{4}$ and $\alpha=\frac{3}{2}$}

These backgrounds have been described in Appendix \ref{ap:exact}.
It is useful to orient the radial variable in the same direction in both cases. For the \texttt{EinSV}$_1$ model, with $\alpha = \frac{3}{4}$, set $z = r-1$. The exterior now has the boundary at $z=0$ and the horizon at $z_\mathcal{H}=1$. For the \texttt{EinSV}$_2$ model, with $\alpha = \frac{3}{2}$, set $z = 1-r$. The boundary is again at $z=0$ and the horizon is at $z_\mathcal{H}=\frac{1}{2}$. For both cases we may define the Eddington-Finkelstein coordinate
\begin{equation}
 v \equiv t-\int_0^z\frac{d z'}{g(z')} \,.
 \label{eq:exact-v}
\end{equation}
In a slight abuse of notation here, $g(z)$ means $g(r(z))$. The metric (\ref{eq:einsv}) then becomes
\begin{equation}
 ds^2=\Omega(z)\left[-g(z) dv^2
 -2\, dv\, dz + dx_1^2+ dx_2^2\right] \,.
 \label{eq:exact-ef-metric}
\end{equation}

As for the previous Schwarzschild case, we write the probe massless scalar field as
\begin{equation}
 \Phi(v,z,x_1)=e^{-i \omega v+ i k x_1}\,z^3 y(z) \,.
 \label{eq:exact-u3-ansatz}
\end{equation}
The wave equation (\ref{eq:wave}), with $m^2=0$, then becomes
\begin{equation}
 g y''+
 \left[g'+g\left(\frac{\Omega'}{\Omega}+\frac{6}{z}\right)
       +2 i \omega\right]y' +\left[
 \frac{3 g}{z}\left(\frac{g'}{g}+ \frac{\Omega'}{\Omega} + \frac{2}{z}\right)
 +i \omega \left(\frac{\Omega'}{\Omega}+\frac{6}{z} \right)
 -q\right]y = 0 \,.
 \label{eq:exact-Y-wave}
\end{equation}
The series expansions at the horizon and at the boundary are then
\be\label{eq:exp2}
y_B(z) = \sum_{j=0}^{j_\text{max}} b_j z^j \,, \qquad y_H(z)=\sum_{j=0}^{j_\text{max}}c_j (z_\mathcal{H}-z)^j \,.
\ee
As previously, $b_0 = c_0 = 1$ and the higher order coefficients are determined recursively by the wave equation (\ref{eq:exact-Y-wave}). The horizon radii $z_\mathcal{H}$ were given above for the two cases.

The natural point at which to calculate the Wronskian of the two series solutions in (\ref{eq:exp2}) is the midpoint $\frac{z_\mathcal{H}}{2}$. It is easily verified, by computing the roots of $g(z)$ and the poles of $\frac{\Omega'}{\Omega}$, that the midpoint is within the radius of convergence of both series. The rest of the analysis then proceeds identically to the Schwarzschild case.

\subsection*{Einstein-Maxwell-dilaton background}

The background is given in Appendix \ref{ap:exact}. Introduce the Eddington-Finkelstein coordinate
\be
v \equiv t - \int_0^r \frac{H(r')^{2/5}}{G(r')} dr' \,,
\ee
so that the metric becomes
\begin{equation}
 ds^2=\frac{H^{4/15}}{r^2}
 \left[-\frac{G d v^2}{H^{4/5}}
 -\frac{2 dv d r}{H^{2/5}} 
 + dx_1^2 + dx_2^2 + dx_3^2\right].
 \label{eq:transition-metric-ef}
\end{equation}
We again write the probe massless scalar field as (with a different power for five dimensions)
\begin{equation}
 \Phi(v,r,x_1)=e^{-i \omega v+ i k x_1}\,r^4 y(r) \,.
 \label{eq:transition-u4-ansatz}
\end{equation} 
The wave equation (\ref{eq:wave}), with $m^2=0$, then becomes
\be
G y'' + \left(\frac{5 G}{r} + G' + 2 i \omega H^{2/5}\right) y' + \left[\frac{4 G'}{r} + 2 i 
\omega H^{2/5} \left(\frac{5}{2 r} + \frac{H'}{5 H} \right)  - q \right] y = 0 \,.
\ee
At large $Q$ the radius of convergence of the power series solution about the $r=0$ boundary goes to zero, due to the zeros of $G$, and the midpoint matching doesn't work. However, for all of the values $Q \lesssim 0.2$ used in Fig.~\ref{fig:transition} the series expansions at the horizon and at the boundary do converge at the midpoint $r=\frac{1}{2}$. We may then follow the same methodology as described for the other backgrounds above.

The quasinormal modes for larger values of $Q$ can be accessed through a change of variables that moves the zeros of $G$ away from the endpoints at $r=0$ and $r=1$. For example, larger (but not arbitrarily large) values of $Q$ can be accessed by setting $\bar r \equiv (1+Q r^2)^{-1/5}$ and then mapping the range back to the interval $[0,1]$ with the normalised coordinate $\hat r = \frac{\bar r-(1+Q)^{-1/5}}{1-(1+Q)^{-1/5}}$. In fact, we used this improved coordinate to compute the quasinormal modes shown in Fig.~\ref{fig:transition}. As previously, accuracy was checked by using different truncation orders in the series expansion and by varying the matching point. We also checked the low-lying modes using the original $r$ coordinate and transporting the field and its derivative from the endpoints to a common matching point using a sequence of overlapping series expansions, of the kind we will describe below.

\subsection*{Einstein-scalar Kasner}

In this case the background is not known analytically. The strategy remains to avoid integration of the Einstein equations and to reduce the problem to a set of polynomial recursion relations.

Recall that the action was given in (\ref{eq:ES}), the background in (\ref{eq:back}) and the equations of motion in Appendix \ref{ap:cor}.
The AdS boundary, horizon and singularity are at
$r=0$, $r=1$ and $r=\infty$, respectively.  We consider a particular numerical background with $\phi(1)=\frac12$ and $\chi(1)\approx 0.048$ (the value of $\chi$ at the horizon is fixed by the requirement that $\chi \to 0$ at the boundary). The Schwarzschild solution is also a solution to the equations of motion, but with $\phi(1)=0$. Using the numerical method described in \cite{Frenkel:2020ysx}, or the method described below, integrating the equations towards the singularity with this horizon data produces the Kasner data shown in the second row of Table \ref{tab:models}.

Before solving the wave equation, we must construct the background (\ref{eq:back}) in a convenient way. We will obtain the exterior on a logarithmic mesh between cutoffs $r_H \approx 1$, close to the horizon, and $r_B \approx 0$, close to the boundary. Specifically, setting $u = \log r$, we will obtain the background fields at values $u_i$ obeying
\begin{equation}
 u_H=u_0 > u_1 > \cdots > u_{i_\text{max}} = u_B \,.                   \label{eq:mesh-order}
\end{equation}
The exact values of $u_i$ used to build the mesh will be described below. The background at $u_H$ is obtained using the regular series expansion of the equations at the horizon. This is uniquely fixed given $\phi(1)$ and $\chi(1)$,
\begin{equation}
 (f,\chi,\phi)(r_H)=\sum_{\ell=0}^{H_{\rm bg}}
 (f^H_\ell,\chi^H_\ell,\phi^H_\ell) (1-r_H)^\ell \,.
 \label{eq:horizon-background-series}
\end{equation}

To propagate the near-horizon values in (\ref{eq:horizon-background-series}) towards the boundary it is useful to work with first order equations. Setting $\dot x \equiv \frac{dx}{du}$ for any variable $x$ and introducing $\pi \equiv \dot \phi$, the background equations can be written in the first order form 
\begin{equation}
 \dot f=3f+\frac14f\pi^2-\frac12(6+\phi^2),\qquad
 \dot\chi=\frac12\pi^2,\qquad
 \dot \phi=\pi,\qquad
 \dot\pi=\frac{-2\phi+\frac12(6+\phi^2)\pi}{f} \,. \label{eq:logr}
\end{equation}
For $u$ close to the node $u_i$ we may use these equations to obtain a power series
\begin{equation}
 (f,\chi,\phi,\pi)(u)=\sum_{\ell=0}^{N_{\rm ch}}
 (f_{i\ell},\chi_{i\ell},\phi_{i\ell},\pi_{i\ell})(u - u_i)^\ell \,.      \label{eq:background-chart}
\end{equation}
By matching a sequence of overlapping power series solutions from node to node, the near-horizon data (\ref{eq:horizon-background-series}) is transported across the mesh into unique near-boundary values of the fields at $r_B$.
Using the boundary data one can compute the source for the scalar field and the temperature of the black hole, as done in \cite{Frenkel:2020ysx} using a more simple numerical integration, and we have checked that the answers agree with the previous results.

We can now compute the quasinormal modes. In 
Eddington--Finkelstein coordinates, the metric is
\begin{equation}
 ds^2=\frac1{r^2}\left[-f\e^{-\chi}d v^2
 -2\e^{-\chi/2}d v\,d r+d x_1^2+d x_2^2\right].              \label{eq:EF-metric}
\end{equation}
The massless scalar field takes the same form (\ref{eq:scalar-ansatz}) as for Schwarzschild. The wave equation is
\be
y'' + \left( \frac{4}{r} - \frac{\chi'}{2} + \frac{f'}{f} + \frac{2 i \omega e^{\chi/2}}{f} \right) y' + \left(\frac{3 f'}{r f} + \frac{4 i \omega e^{\chi/2}}{r f} - \frac{3 \chi'}{2 r} - \frac{q}{f} \right) y = 0 \,.
\ee
The logic now follows as for the previous backgrounds we have discussed. We consider two solutions. The first is defined by a boundary expansion at small $r$
\begin{equation}
 y_B(r)=1-i \omega r + \cdots \,.   
 \label{eq:bseries}
\end{equation}
We use this expansion to obtain $y_B$ and its derivative at $r_B$. For sufficiently small $r_B$, it is sufficient to extend the series only up to quadratic order. All of the coefficients in the series are uniquely determined given values of $\omega, q$ and the background geometry. The second solution is defined by a regular near-horizon expansion
\be
y_H(r) = 1 + \sum_{n=1}^{H_{\rm mode}} c_n (1-r)^n \,.
\ee
Again, all of the coefficients are determined by $\omega, q$ and the background geometry.

As for the background, we transport these two solutions across the nodes $u_i$ using overlapping series expansions centred on each node. For this process, it is convenient to introduce a normalisation-invariant combination
\be
R \equiv {\dot{y} \over y} \,,
\ee
so that near every node
\be
R(u) = \sum_{n=0}^{N_{\text{ch}}-1} r_{in}(u-u_i)^n \,.
\ee
We transport both solutions to a common point $u_m$ and evaluate the normalised Wronskian
\be
W(\omega,q) \equiv {y_H \dot y_B - \dot y_H y_B \over y_H y_B}  =R_B(u_m)-R_H(u_m) \,.
\ee
The search for the frequencies at which the Wronskian vanishes and the subsequent computation of the coefficient $D_n$ now proceed exactly as for the Schwarzschild case discussed above.

Table \ref{tab:scheduled-orders} shows the numerical parameters that we used in two independent runs (`C' and `D'). Numerical stability was checked by comparing runs C and D for every mode, for both $\omega_n$ and $D_n$.
The table shows the cutoffs $(H_{\rm bg},H_{\rm mode},N_{\rm ch})$ in the series expansions, the geometrical cutoffs and matching point $(1-r_H, r_B, r_m)$ and the decimal working precision `prec'. The Newton tolerances were $10^{-44}$ for run C and $10^{-52}$ for run D, with at most 16 iterations. 
\begin{table}[h]
\centering
\small
\begin{tabular}{@{}ccccccccc@{}}
\toprule
$n$ & $\text{run}$ & $\text{prec}$ & $H_{\rm bg}$ & $H_{\rm mode}$
& $N_{\rm ch}$ & $1-r_H$ & $r_B$ & $r_m$\\
\midrule
$1{:}12$  & C & $150$     & $175$     & $155$     & $68$     & $0.028$ & $10^{-14}$ & $0.47$\\
$1{:}12$  & D & $220$     & $215$     & $190$     & $76$     & $0.021$ & $10^{-17}$ & $0.59$\\
$13{:}20$ & C & $120$     & $155$     & $135$     & $56$     & $0.028$ & $10^{-14}$ & $0.47$\\
$13{:}20$ & D & $180$     & $185$     & $160$     & $64$     & $0.021$ & $10^{-17}$ & $0.59$\\
$21{:}40$ & C & $40+4n$  & $95+3n$  & $75+3n$  & $16+2n$  & $0.028$ & $10^{-14}$ & $0.47$\\
$21{:}40$ & D & $100+4n$  & $125+3n$  & $100+3n$  & $24+2n$  & $0.021$ & $10^{-17}$ & $0.59$\\
\bottomrule
\end{tabular}
\caption{Numerical parameters used in two computations of the quasinormal modes.} 
\label{tab:scheduled-orders}
\end{table}
The parameters for run D were chosen, among other reasons, to ensure that the last retained term of every series expansion was less than $10^{-16}$ times the value of the polynomial. We also performed additional checks on the results for $D_n$ by explicitly computing some modes at two nearby values of $q$ and taking the derivative.

Finally, we can note the meshes used. The mesh was finer closer to the horizon. The radial direction was split up into segments: $r_H \to r_m \to r_t \to r_B$, where $r_t = 0.05$. In each segment we used a uniform spacing $\Delta u$, whose values are shown in Table \ref{tab:mesh}.
\begin{table}[h]
\centering
\begin{tabular}{@{}ccccc@{}}
\toprule
mode range & run
& $\substack{|\Delta u|\\ r_H\to r_m}$
& $\substack{|\Delta u|\\ r_m\to r_t}$
& $\substack{|\Delta u|\\ r_t\to r_B}$ \\
\midrule
$n\leq 12$ & C & $0.072662$ & $0.074690$ & $0.495601$ \\
$n\leq 12$ & D & $0.056268$ & $0.057398$ & $0.415497$  \\
$n\geq 13$ & C & $0.080736$ & $0.089628$ & $0.596744$  \\
$n\geq 13$ & D & $0.063301$ & $0.068558$ & $0.495181$ \\
\bottomrule
\end{tabular}
\caption{Spacings used for the meshes in the two runs.}
\label{tab:mesh}
\end{table}

\section{First-order perturbative computation}
\label{app:first-order-monodromy}

This appendix provides more details for the steps going from the integral expression (\ref{eq:l2}) for the perturbation  to the large-$x$ result (\ref{eq:shift}). We assume that $0<\alpha<1$, which guarantees that the integrals below converge.

On the branch used in \S\ref{sec:qnm2}, the required Hankel continuation formulae are
\be
 H_0^{(1)}(x)=2H_0^{(1)}(e^{i\pi}x)+H_0^{(2)}(e^{i\pi}x) \,, \qquad H_0^{(2)}(x) =-H_0^{(1)}(e^{i\pi}x) \,.
\ee
Applying these identities after the change of variables $t=e^{-i\pi}s$ gives 
\be
 H_0^{(1)}(e^{-i\pi}s) =2H_0^{(1)}(s)+H_0^{(2)}(s) \,, \qquad H_0^{(2)}(e^{-i\pi}s) =-H_0^{(1)}(s) \,.
 \label{eq:first-mon-continuation}
\ee
The second line of (\ref{eq:l2}) becomes, including the change in measure
$t^{\a-1}dt = e^{-i\pi\a}s^{\a-1}ds$, 
\begin{align}
 e^{-i\pi\a}\int_0^{e^{i\pi}x} \,s^{\a-1}
 \Big(2H_0^{(1)}(s)+H_0^{(2)}(s)\Big)
 \Big(H_0^{(1)}(e^{i\pi}x)H_0^{(2)}(s)
      -H_0^{(2)}(e^{i\pi}x)H_0^{(1)}(s)\Big)ds.
 \label{eq:first-mon-rotated-integral}
\end{align}
For the large-$e^{i\pi}x$ limit relevant to the monodromy, the upper limit in
(\ref{eq:first-mon-rotated-integral}) may be sent to infinity.  The integrals that arise are then
\be
 \int_0^\infty s^{\a-1}H_0^{(1)}(s)H_0^{(2)}(s) ds
 =\frac{\Gamma(\frac{\a}{2})^3
          \Gamma(\frac{1-\a}{2})}{\pi^{5/2}} \,,
\ee
and
\begin{align}
\int_0^\infty s^{\a-1}\left[H_0^{(1)}(s)\right]^2 ds & =
e^{i \pi \a} \int_0^\infty s^{\a-1}\left[H_0^{(2)}(s)\right]^2 d s \nonumber \\
 & =-e^{i\frac{\pi}{2}\a}
 \frac{\Gamma(\frac{\a}{2})^3}
      {\pi^{3/2}\Gamma(\frac{1+\a}{2})} = -e^{i\frac{\pi}{2}\a} \frac{\Gamma(\frac{\a}{2})^3
          \Gamma(\frac{1-\a}{2})}{\pi^{5/2}}
   \cos\!\left(\frac{\pi\a}{2}\right)\,.
\end{align}
In the final step we used the gamma function reflection formula.

We now substitute the above integral expressions into (\ref{eq:l2}), collect the Hankel functions, factor out the common gamma function coefficients and use the two trigonometric identities
\begin{align}
 \cos\!\left(\frac{\pi\a}{2}\right)
 \left(e^{i\frac{\pi}{2}\a}-e^{-i\frac{3\pi}{2}\a}\right)
 +2\left(e^{-i\pi\a}-1\right) & =-4i e^{-i\frac{\pi}{2}\a}
 \sin^3\!\left(\frac{\pi\a}{2}\right) \,, \\
 2e^{-i\frac{\pi}{2}\a}\cos\!\left(\frac{\pi\a}{2}\right) -(1+e^{-i\pi\a})
& = 0 \,.
\end{align} 
This gives 
\begin{align}
 \delta\psi(x)
 &\sim \frac{i\pi}{4}
 \frac{\Gamma(\frac{\a}{2})^3
       \Gamma(\frac{1-\a}{2})}{\pi^{5/2}}
 \Bigg[
 -4i e^{-i\frac{\pi}{2}\a}\sin^3\!\left(\frac{\pi\a}{2}\right)
 H_0^{(1)}(e^{i\pi}x) + 0 \times H_0^{(2)}(e^{i\pi}x)
 \Bigg] \nonumber\\
 &=\frac{e^{-i\frac{\pi}{2}\a}\pi^{3/2}
          \Gamma(\frac{1-\a}{2})}
         {\Gamma(1-\frac{\a}{2})^3}
 H_0^{(1)}(e^{i\pi}x),
 \label{eq:first-mon-final}
\end{align}
where in the second line we used
$\Gamma(\frac{\a}{2})\sin(\frac{\pi\a}{2})=\pi/\Gamma(1-\frac{\a}{2})$.

\section{Second-order perturbative computation}
\label{app:analytic-alpha}

In this appendix we derive \eqref{eq:leadcorrectionsalpha}. Equations~\eqref{eq:p1} and \eqref{eq:p2} can both be written as
\begin{equation}
 \psi''+\frac{1}{x}\psi'+\psi
 =-\left(\frac{\epsilon_1}{x^\beta}
 +\frac{\epsilon_2}{x^{2-\alpha}}\right)\psi \,.
 \label{eq:appendix-F-master}
\end{equation}
We expand the solution that is infalling on the positive $x$ axis as
\begin{equation}
 \psi=H_0^{(1)}(x)
 +\epsilon_1\delta_{10}\psi
 +\epsilon_2\delta_{01}\psi
 +\epsilon_1\epsilon_2\delta_{11}\psi+\cdots \,.
 \label{eq:appendix-F-expansion}
\end{equation}
Here $\delta_{10}\psi$ and $\delta_{01}\psi$ are the first-order solutions
\eqref{eq:l1} with $\alpha$ there replaced by $2-\beta$ and $\alpha$,
respectively. The contribution of such terms to the monodromy has been obtained in Appendix~\ref{app:first-order-monodromy}. We wish to obtain the leading correction to the dispersion $\frac{\pa \omega_n}{\pa k^2}$ due to the change in the geometry. For that we need to compute the monodromy due to the new, mixed term in (\ref{eq:appendix-F-expansion}).

A convenient particular solution for the mixed term is
\begin{align}
 \delta_{11}\psi(x)
 =\frac{i\pi}{4}\Bigg\{&
 -H_0^{(1)}(x)\int_x^\infty \ H_0^{(2)}(t)
 \left[t^{1-\beta}\delta_{01}\psi(t)
       +t^{\alpha-1}\delta_{10}\psi(t)\right] dt \nonumber\\
 &+H_0^{(2)}(x)\int_x^\infty \ H_0^{(1)}(t)
 \left[t^{1-\beta}\delta_{01}\psi(t)
       +t^{\alpha-1}\delta_{10}\psi(t)\right]dt\Bigg\}.
 \label{eq:appendix-F-mixed-solution}
\end{align}
This follows from the same Green's function construction as
Eq.~\eqref{eq:l1} in the main text. Again, an arbitrary multiple of $H_0^{(1)}(x)$ may be added but does not affect the monodromy. The integrals in \eqref{eq:appendix-F-mixed-solution} are convergent in the strip
$0<\text{Re}(2-\beta),\operatorname{Re}\alpha <1$. These conditions fail for the values $\beta = \frac{2}{3}$ and $-\frac{10}{11}$ needed below, and we must therefore analytically continue the integrals. The analytic continuation can be done numerically as follows. Firstly, regularise the integral with a large $t$ cutoff $T$. Secondly, subtract any terms that diverge as $T \to \infty$.
Finally, remove the cutoff to obtain a finite answer. Because the subtracted terms vanish for values of $\a,\beta$ where the original integral is convergent, they have no effect there and thereby provide an analytic continuation of the integral.

To compute the monodromy we express \eqref{eq:appendix-F-mixed-solution} as a function of $e^{i\pi}x$, exactly as in Appendix~\ref{app:first-order-monodromy}. In the present case, however, the coefficient of $H_0^{(2)}(e^{i\pi}x)$ obtained from \eqref{eq:appendix-F-mixed-solution} does not vanish. We can remove this term with a shift of \eqref{eq:appendix-F-mixed-solution} by $cH_0^{(1)}(x)$, which leads to a corresponding shift of the rotated expression by $2 c H_0^{(1)}(e^{i\pi}x)+c H_0^{(2)}(e^{i\pi}x)$. We may then choose $c$ to cancel the $H_0^{(2)}(e^{i\pi}x)$ term generated from \eqref{eq:appendix-F-mixed-solution}, so that
as $e^{i\pi}x\rightarrow+\infty$ we obtain
\begin{equation}
 \delta_{11}\psi(x)
 \sim {\cal C}_{2-\beta,\alpha}
 H_0^{(1)}(e^{i\pi}x) \,,
 \label{eq:appendix-F-mixed-definition}
\end{equation}
where numerical evaluation of the integrals, analytically continued as described above, gives
\begin{align}
 {\cal C}_{\frac{4}{3},\,\frac{2}{3}}
 &\approx 16.324 \,,
 \label{eq:appendix-F-small-mixed}\\
 {\cal C}_{\frac{32}{11},\,\frac{10}{11}}
 &\approx 35.376+10.387 i \,.
 \label{eq:appendix-F-large-mixed}
\end{align}
The first number here is numerically consistent with $3\sqrt3\,\pi$.

It remains to show how the monodromy shifts (\ref{eq:appendix-F-small-mixed}) and (\ref{eq:appendix-F-large-mixed}) enter the effective exponent $\alpha_n$, defined in (\ref{eq:alpha}).
We must also keep track of the first-order effects. For convenience,
denote the first-order coefficient derived in
Appendix~\ref{app:first-order-monodromy} by
\begin{equation}
 {\cal A}_\a \equiv
 \frac{e^{-i\frac{\pi}{2} \a}\pi^{3/2}\Gamma(\frac{1-\a}{2})}
      {\Gamma(1-\frac{\a}{2})^3} \,.
 \label{eq:appendix-F-Ap}
\end{equation}
Exactly as in the discussion above (\ref{eq:wfull}) in the main text, the quasinormal mode quantisation condition (\ref{eq:cond}) is shifted by $2 \to 2 e^{\frac{1}{2} X}$, where now
\be
X = \cdots+\epsilon_1\epsilon_2
 \left[{\cal C}_{2-\beta,\alpha}
 -\frac12{\cal A}_{2-\beta}{\cal A}_{\alpha}\right]+\cdots \,.
 \label{eq:appendix-F-log-ratio}
\ee
The quasinormal frequency is then shifted by $\delta \omega = \frac{i X}{4 {\mathcal Z_+}}$.

Specialising to the case at hand, we are interested in two limits.
From \eqref{eq:p1} and \eqref{eq:p2},
\begin{alignat}{2}
|\hat \omega| \ll 1 \; : \;\; & \epsilon_1=\frac45\frac{1}{3^{2/3}}\hat\omega^{\frac{2}{3}} \,,
 &\;\;&
 \epsilon_2= \frac{1}{3^{4/3}} \frac{\hat k^2}{\hat\omega^{\frac{2}{3}}} \,.
 \label{eq:appendix-F-small-eps} \\
 |\hat \omega| \gg 1 \; : \;\;\; & \epsilon_1=\frac45\left(\frac{11}{5}\right)^{\frac{10}{11}}
 \frac{1}{\hat\omega^{\frac{10}{11}}} \,,
 &\;\;\;&
 \epsilon_2=\left(\frac5{11}\right)^{\frac{12}{11}}
 \frac{\hat k^2}{\hat\omega^{\frac{10}{11}}} \,. 
 \label{eq:appendix-F-large-eps}
\end{alignat}
In these two limits the dispersion becomes, respectively,
\begin{align}
D_n & = \frac{1}{\sqrt{Q}} \left. \frac{\pa \hat \omega_n}{\pa \hat k^2}\right|_{\hat k = 0} = 
\frac{i Q}{4 {\mathcal Z}_+} \left\{
\begin{array}{l}
 \displaystyle \frac{1}{3^{4/3}} \frac{{\cal A}_{2/3}}{\hat\omega^{\frac{2}{3}}}  + \frac4{45}\left[{\cal C}_{\frac{4}{3},\frac{2}{3}}
-\frac12{\cal A}_{\frac{4}{3}}{\cal A}_{\frac{2}{3}}\right]+\cdots \\
\displaystyle \left(\frac5{11}\right)^{\frac{12}{11}}
 \frac{{\cal A}_{10/11}}{\hat\omega^{\frac{10}{11}}} + \frac45\left(\frac5{11}\right)^{2/11}
\left[{\cal C}_{\frac{32}{11},\frac{10}{11}}
-\frac12{\cal A}_{\frac{32}{11}}{\cal A}_{\frac{10}{11}}\right]
\frac{1}{\hat\omega^{\frac{20}{11}}}+\cdots
\end{array} \right. \nonumber \\
& \equiv D(\hat \omega) \,. \label{eq:appendix-F-dispersion}
\end{align}

In the scaling limit (\ref{eq:scalinglimit}) where $Q \to 0$ we have,
using \eqref{eq:modes1}, \eqref{eq:defxin} and \eqref{eq:NQ}, that for the correction to the $n$th overtone
\be
\hat \omega = \hat \omega_{0n} = Q^{3/2} \omega_{0n} \approx e^{- i \frac{\pi}{4}} \xi_n \,.
\ee
Here we used the fact that as $Q \to 0$ the optical length ${\mathcal Z}_+$ becomes (as noted below (\ref{eq:NQ}) in the main text) that of the Schwarzschild-AdS$_5$ geometry, which has $\arg{\cal Z}_+ = \frac{\pi}{4}$. In the scaling limit we can evaluate the effective exponent (\ref{eq:alpha}) using the continuum expression given below (\ref{eq:alpha}), so that
\be\label{eq:f3}
\alpha_n \approx - \text{Re}\, \frac{2 n \pa_n D_n + n^2 \pa_n^2 D_n}{n \pa_n D_n + D_n} \approx \left. - \text{Re}\, \frac{2 \hat \omega D'(\hat \omega) + \hat \omega^2 D''(\hat \omega)}{\hat \omega D'(\hat \omega) + D(\hat \omega)} \right|_{\hat \omega = e^{- i \frac{\pi}{4}} \xi_n} \,.
\ee
The overall factor of $\frac{iQ}{4{\cal Z}_+}$ in (\ref{eq:appendix-F-dispersion}) cancels in this expression.

It is now just a question of substituting (\ref{eq:appendix-F-dispersion}) into (\ref{eq:f3}). For $|\hat\omega|\ll1$ this gives
\begin{equation}
\alpha_n = \frac23
-\frac{8}{5\,3^{2/3}}\operatorname{Re}\!\left[
e^{-i\frac{\pi}{6}}\left(
\frac{{\cal C}_{\frac{4}{3},\frac{2}{3}}}{{\cal A}_{\frac{2}{3}}}
-\frac12{\cal A}_{\frac{4}{3}}\right)\right]\xi_n^{2/3}
+o(\xi_n^{2/3}),
\qquad \xi_n\to0 \,.
\label{eq:appendix-F-small-curve}
\end{equation}
For $|\hat\omega|\gg1$ we obtain
\begin{equation}
\alpha_n={}\frac{10}{11}
-\frac{72}{11}\left(\frac{11}{5}\right)^{\frac{10}{11}}
\operatorname{Re}\!\left[
e^{\frac{5}{22}i \pi}\left(
\frac{{\cal C}_{\frac{32}{11},\frac{10}{11}}}{{\cal A}_{\frac{10}{11}}}
-\frac12{\cal A}_{\frac{32}{11}}\right)\right] \frac{1}{\xi_n^{10/11}}
+o(\xi_n^{-10/11}),\qquad \xi_n\to\infty \,.
\label{eq:appendix-F-large-curve}
\end{equation}
Finally, evaluating ${\cal A}$ using \eqref{eq:appendix-F-Ap} and using
the numerical values (to a higher accuracy than quoted) of ${\cal C}$ given in
\eqref{eq:appendix-F-small-mixed} and \eqref{eq:appendix-F-large-mixed}, we obtain 
\begin{equation}
\begin{aligned}
\alpha_n &=\frac23-0.21778\,\xi_n^{2/3}+\cdots,
&\qquad \xi_n\to0,\\
\alpha_n &=\frac{10}{11}- \frac{0.27213}{\xi_n^{10/11}}+\cdots,
&\qquad \xi_n\to\infty \,.
\end{aligned}
\label{eq:appendix-F-numerical-curves}
\end{equation}
These are the endpoint corrections quoted in
Eq.~\eqref{eq:leadcorrectionsalpha}.

\bibliographystyle{ourbst}
\bibliography{refs}

\end{document}